\documentclass[reprint, aps, superscriptaddress, citeautoscript, floatfix, amsmath, amssymb, prb]{revtex4-1}

\usepackage{lmodern,microtype}
\usepackage[mathlines]{lineno}
\relax

\usepackage{lipsum}
\usepackage{graphicx}
\usepackage{color,soul}
\usepackage[dvipsnames]{xcolor}
\usepackage[caption=false]{subfig}
\usepackage{array,booktabs,paralist,dcolumn}
\usepackage{enumitem}
\usepackage{multirow}
\usepackage{mathtools,amssymb}
\usepackage{mathdesign,bm}

\usepackage{hyperref}
\usepackage{filecontents}

\setlist[itemize]{leftmargin=*}

\begin{document}
\title{Temperature-dependent optical and magneto-optical spectra of ferromagnetic BCC Fe}

\author{Kisung Kang}
\affiliation{ 
The NOMAD Laboratory at the FHI of the Max-Planck-Gesellschaft and IRIS-Adlershof of the Humboldt-Universit\"{a}t zu Berlin, Faradayweg 4-6, 14195 Berlin, Germany
}

\author{David G. Cahill}
\affiliation{Department of Materials Science and Engineering, University of Illinois at Urbana-Champaign, Urbana, IL 61801, USA}
\affiliation{Materials Research Laboratory, University of Illinois at Urbana-Champaign, Urbana, IL 61801, USA}

\author{Andr\'e Schleife}
\email{schleife@illinois.edu}
\affiliation{Department of Materials Science and Engineering, University of Illinois at Urbana-Champaign, Urbana, IL 61801, USA}
\affiliation{Materials Research Laboratory, University of Illinois at Urbana-Champaign, Urbana, IL 61801, USA}
\affiliation{National Center for Supercomputing Applications, University of Illinois at Urbana-Champaign, Urbana, IL 61801, USA}

\begin{abstract}
Optical and magneto-optical properties of magnetic materials have been widely exploited to characterize magnetic structures and phenomena, however, their temperature dependence is not well understood.
This study implements the supercell approach with thermal lattice and magnetic disorders to obtain optical and magneto-optical spectra at finite temperatures based on Williams-Lax theory.
Our results show that large optical spectrum signals are generated at photon energies below 1\,eV, originating from the phonon- and magnon-assisted intraband transitions as lattice and magnetic temperatures increase.
In addition, the prominent peak near 2.7\,eV is redshifted proportionally to magnetic temperature but depends much less on lattice temperature.
By analyzing unfolded bands, we show that the reduction of exchange splitting due to the thermal demagnetization causes this redshift.
Our unfolded electronic band structure with magnetic disorder shows band kinks, which are characteristic evidence of the coupling between electrons and magnetic excitations.
First-order magneto-optical spectra at finite temperature are also predicted, but due to their small magnitude suffer more from sampling errors.
We discuss the effect of zero-point vibrations and the connection of these simulations to the Drude model for intraband transitions.
\end{abstract}

\maketitle

\section{\label{sec:intro}Introduction}

Optical and magneto-optical spectra play a crucial role in revealing information about magnetic properties and the electronic structure of materials.
The temperature dependence of such excited-state properties of magnetic materials encodes and provides access to the intertwined electron, lattice, and magnetic excitations \cite{Sugano:2000, Siddiqui:2020}.
However, acquiring detailed information about the individual effects remains challenging due to the overlapping energy scale and the couplings among electrons, phonons, and magnons.
Therefore, there is a demand for a systematic investigation into how electron, lattice, and magnetic temperatures are linked to optical and magneto-optical spectra.
Such an investigation can provide direct insight into thermal effects in magnetic materials.

In experiments, (magneto-)optical spectra at finite temperatures have been popularly measured to investigate the magnetic properties of materials both in and out of equilibrium.
Static measurements of magneto-optical signals at different temperatures provide insights into equilibrium properties.
This approach can be employed to indirectly visualize how magnetic structure changes as temperature increases, allowing for the determination of the magnetic phase transition and its critical temperature of magnetic materials\cite{Hansen:1989, Kalashnikova:2005, Balk:2019, Saidl:2016}.
Furthermore, pump-probe time-resolved (TR) magneto-optical Kerr effect (MOKE) measurements have been used to uncover the physical origin of magnetic properties out of equilibrium.
This method enables, for instance, the identification of temperature-dependent magnon frequencies and magnon relaxation rates, concluding the importance of the biquadratic exchange interaction for the temperature dependence in antiferromagnetic NiO\cite{Kohmoto:2018}.
In antiferromagnetic Fe$_2$As, the temperature gradient of magnetic birefringence from TR-MOKE measurements shows a similar temperature trend as the magnetic heat capacity\cite{Yang:2019}.
Since these quantities are connected via exchange interactions, it is possible to determine the N\'eel temperature through optical response measurements \cite{Yang:2019}.
However, unveiling the underlying physical origin of temperature-dependent optical spectra via experiments remains challenging due to the intricate manipulation of individual electron, lattice, and magnetic temperatures being a sophisticated task.
Achieving this by using first-principles electronic-structure simulations is the main goal of this paper.

Theories regarding the temperature-dependent electronic structures and optical properties of materials have been developed since the 1950s, motivated by the need to explain various experimental investigations\cite{Williams:1951, Lax:1952, Hall:1954, Allen:1976}.
These theories are applicable to non-magnetic materials by incorporating lattice temperatures described by nuclear vibrational motions.
Firstly, the Williams and Lax (WL) theory is based on a semiclassical approximation, employing an ensemble average over atomic structures perturbed by thermal atomic vibrations.
This approach can elucidate changes in both the electronic eigenvalues and optical transition matrix elements with varying temperature\cite{Williams:1951, Lax:1952}.
The Hall, Bardeen, and Blatt (HBB) theory describes temperature-dependent optical spectra by employing varying optical transition matrix elements but does not incorporate thermal changes in electronic eigenvalues\cite{Hall:1954}.
This theory is grounded in perturbation theory using electron-phonon matrix elements and a ground-state electronic structure\cite{Hall:1954}.
Lastly, the Allen-Heine (AH) theory explains temperature-dependent changes in the electronic structures due to electron-phonon coupling\cite{Allen:1976}.
These theories serve as the fundamental theoretical background for modern computational methods aiming to predict the electronic structure and optical properties of non-magnetic materials at finite temperatures\cite{Giustino:2017, Zacharias:2015}.

Computational approaches to study electronic and optical properties of \emph{magnetic materials} at finite temperatures are still in a nascent stage and most studies have focused on simple \emph{non-magnetic materials} with pure elements or simple semiconductors \cite{Kaufmann:1998, Noffsinger:2012, Patrick:2014, Zacharias:2015, Peelaers:2015, Zacharias:2016, Brown:2016, Xu:2017, Kang:2018, Monserrat:2018, Morris:2018, Bravic:2019, Liu:2021, Huang:2021}.
Recently developed methods for computing phonon-assisted optical properties at finite temperature have been applied to various non-magnetic materials, including semiconductors \cite{Noffsinger:2012, Patrick:2014, Zacharias:2015, Zacharias:2016, Kang:2018, Monserrat:2018, Bravic:2019}, metals\cite{Kaufmann:1998, Brown:2016, Xu:2017, Liu:2021}, transparent conducting oxides\cite{Peelaers:2015, Morris:2018}, and 2D semiconductors\cite{Huang:2021}.
With the assistance of phonons, indirect interband transitions can be addressed, which captures the onset of the dielectric response correctly as a function of the photon energy in semiconductors\cite{Noffsinger:2012, Patrick:2014}.
This has been a critical discrepancy between experiments and first-principles calculations e.g.\ for bulk silicon\cite{Noffsinger:2012}, perovskite BaSnO$_3$\cite{Kang:2018, Monserrat:2018}, and BAs\cite{Bravic:2019}.
In addition, in doped systems, such as doped SnO$_{2}$\cite{Peelaers:2015}, as well as metals, phonon-assisted intraband transitions occur.
In a first-principles study of frequency-dependent thermoreflectance spectra of BCC Au it was shown that the phonon temperature dominates the temperature dependence of optical spectra\cite{Liu:2021}.
These studies of the impact of finite lattice temperatures on optical spectra indicate that such thermal effects might play a role also in magnetic materials.

Historically, computational work has focused on simulating reliable optical spectra at 0\,K and describing the electronic structure at finite temperatures through first-principles approaches.
Optical and magneto-optical spectra of magnetic materials without any thermal effect\cite{Oppeneer:1992, Werner:2009} have been well investigated using electronic band structures from density functional theory (DFT).
The temperature-dependent electronic band structures of magnetic semiconductors and metals are studied by combining molecular dynamics and atomistic spin dynamics\cite{Gambino:2022} and the Green's function ($GW+T$) method, which includes electron-magnon scattering\cite{Muller:2019, Mylnczak:2019}, respectively.
However, temperature-dependent (magneto-)optical properties of \emph{magnetic materials} have been scarcely investigated.

At low temperatures, the (magneto-)optical properties of ferromagnetic BCC Fe have been investigated using various computational methods, because it is among the most well-investigated materials regarding its optical properties in experiments\cite{Yolken:1965, Blotin:1969, Johnson:1974, Siddiqui:1977, Ordal:1985, Ordal:1988, Silber:2019, Weaver:1979, Ferguson:1969, Krinchik:1968}.
Particularly, optical and magneto-optical spectra of ferromagnetic BCC Fe at 0\,K have been studied using DFT \cite{Singh:1975, Oppeneer:1992, Miyazawa:1999, Werner:2009, Stejskal:2018, Silber:2019, Stejskal:2021}.
These investigations yield good overall agreement when compared with measured spectra\cite{Yolken:1965, Blotin:1969, Johnson:1974, Siddiqui:1977, Ordal:1985, Ordal:1988, Silber:2019, Weaver:1979, Ferguson:1969, Krinchik:1968}, but struggle to capture intraband transitions.
Describing electron-electron interactions
in ferromagnetic BCC Fe at 0\,K through the $GW$ approximation\cite{Hedin:1965} resulted in more accurate band structures with a correction of the overestimated bandwidth of 3\,$d$ valence states by DFT-LSDA, however, a discrepancy for exchange splitting and magnetic moments remains\cite{Yamasaki:2003, Okumura:2019}.

At finite temperatures, the electronic band structure of ferromagnetic BCC Fe has been indirectly examined employing advanced computational methods that include scattering effects.
Scattering effects are one of the main contributors to the thermal change of (magneto-)optical spectra, alongside electronic occupations and thermal expansion\cite{Wilson:2012}.
In magnetic materials, such effects encompass contributions from impurities, electron-electron coupling, electron-phonon coupling, and electron-magnon coupling.
Scattering effects from impurities were addressed by dynamical mean-field theory (DMFT) via the Anderson impurity model, which can describe the thermal scattering effects from impurities\cite{Georges:1996, Katsnelson:1999, Barriga:2009, Grechnev:2007}.
DMFT introduces the quantum many-body system to an impurity model with the correlation and scattering effects and uses the Hubbard model with free parameters to describe the electron-electron scattering\cite{Georges:1996}.
DFT+DMFT successfully captures the quasiparticle damping of energy states below the Fermi level on the electronic structure of ferromagnetic BCC Fe\cite{Katsnelson:1999, Barriga:2009, Grechnev:2007}.
The overall renormalization of the electronic structure shows good agreement with photoemission measurements\cite{Katsnelson:1999, Barriga:2009, Grechnev:2007}.

Scattering effects from electron-magnon coupling are associated with magnetic temperature, and thus the consideration of these effects can indirectly extract insight into how the electronic structure changes at finite magnetic temperatures.
The effect of electron-magnon scattering on the electronic band structure of ferromagnetic BCC Fe has been recently studied via $GT$ self-energy calculations that are analogous to the $GW$ approximation for electron-electron scattering \cite{Muller:2019, Mylnczak:2019}.
Band structure renormalization with electron-magnon scattering can describe the band anomalies caused by spin-wave excitations and Stoner excitations, which are observed in photoemission experiments but not explained by DMFT studies\cite{Muller:2019, Mylnczak:2019}.
The discrepancies between the measured spectrum and the $GT$ method, including the overestimated damping effect, $d$ bandwidth, and band anomalies, are recently solved by the combination of the $GW$ approximation with $GT$ self energies\cite{Nabok:2021}.
These methods to reliably model the electronic structure of magnetic materials are very recent, and thus their expansion to describe optical spectra has not been implemented yet.

Despite earlier mentioned efforts, the implementation of first-principles studies on temperature-dependent optical spectra for magnetic materials has been limited, primarily due to the challenge of simultaneously modeling electron, lattice, and magnetic temperatures.
While the methods mentioned earlier for non-magnetic materials can be employed for lattice temperatures, it is crucial to find an appropriate way to incorporate the magnetic temperature into the approach.
For example, electron-phonon scattering in $\mathbf{q}$-space can be obtained from the electronic band structure, phonon dispersion, and electron-phonon coupling calculations\cite{Ponce:2016}.
The phonon-assisted intraband contributions in materials can be described in HBB theory using perturbative methods\cite{Kioupakis:2010, Peelaers:2015, Brown:2016}, focusing on electron-phonon coupling matrix elements with momentum transfer through calculated electron-phonon coupling coefficients.
Alternatively, an ensemble average of the results obtained from explicit supercell calculations based on WL theory can be computed to encompass the full anharmonicity \cite{Zhang:2018, Zacharias:2020}.
For the inclusion of magnetic temperature, the latter approach is more intuitive, as the magnetic temperature can be incorporated by introducing magnetic fluctuation atop atomic vibration\cite{Gambino:2022}.
In addition, it is imperative to appropriately consider scattering effects from electron-electron couplings.
Frequency-dependent electron-electron scattering can be computed using the $GW$ approach\cite{Hedin:1965, Zhukov:2005, Cazzaniga:2012}.
In the infrared photon energy range, intraband contributions are commonly described using the Drude model and an electron scattering rate that originates from the summation of electron-electron and electron-phonon scattering in non-magnetic materials\cite{Xu:2017}.

In this study, we apply a method based on Williams-Lax theory\cite{Williams:1951, Lax:1952, Zacharias:2015} to investigate temperature-dependent optical and magneto-optical properties of ferromagnetic BCC Fe.
This approach can simultaneously describe electron, lattice, and magnetic temperatures, as explained below, which has not been achieved using perturbative methods.
Finite electronic temperature is described using thermal, Fermi-Dirac distributed occupation numbers within Mermin DFT \cite{Mermin:1965}.
Lattice and magnetic temperatures are described on the same footing within WL theory.
While this theory was developed for nuclear vibrational motion, we are applying it also to magnons, as another bosonic excitation.
Explicit supercell calculations are used to model the corresponding \emph{atomic displacements} using harmonic sampling from the phonon eigenvectors and \emph{magnetic fluctuations} from classical atomistic spin dynamics simulations using magnetic interaction parameters obtained from DFT.
In principle, our approach can fully address electron-phonon and electron-magnon couplings.
Optical and first-order magneto-optical spectra are then calculated within the independent particle approximation and thoroughly compared with measured spectra.
In the low photon energy range, the evidence of phonon and magnon-assisted intraband transitions is captured by our simulations and we explain the physical origin of this phenomenon using optical matrix elements and joint density of states.
In addition, we explain a characteristic red-shift of a peak in the optical spectra near 3\,eV by invoking finite magnetic temperature.

The remainder of this paper is organized as follows:
In Sec.\,\ref{sec:methods} we discuss the detailed methodology for computing temperature-dependent optical spectra.
Computational details are provided in Sec.\,\ref{sec:comp}.
In Sec.\,\ref{sec:results}, we present the calculated optical spectra and first-order magneto-optical spectra of ferromagnetic BCC Fe with electronic, lattice, and magnetic temperatures of 300\,K.
Section \ref{sec:disscusion} presents our analysis of the physical origin of the temperature dependence and Sec.\,\ref{sec:cncl} summarizes and concludes our work.

\section{\label{sec:methods}Theoretical Methods}

Thermal excitations in magnetic materials comprise of electron, phonon, and magnon contributions, all of which can be modeled in first-principles simulations.
We account for finite electronic temperature using Mermin DFT \cite{Mermin:1965}, by means of occupation numbers $n_{i,\mathbf{k}}$ of Kohn-Sham states that follow a Fermi-Dirac distribution,
\begin{equation}
\label{eq:FermiDirac}
n_{i,\mathbf{k}}=\frac{1}{e^{\beta(\epsilon_{i,\mathbf{k}}-E_{\mathrm{F}})}+1},
\end{equation}
where $\beta=1/k_{\mathrm{B}}T$, $k_{\mathrm{B}}$ is the Boltzmann constant, $\epsilon_{i,\mathbf{k}}$ is the Kohn-Sham eigenvalue at band index $i$ and $\mathbf{k}$-point $\mathbf{k}$, and $E_{\mathrm{F}}$ is the Fermi energy.
In this work, we study electronic temperatures between 5\,K and 300\,K.

Lattice and magnetic temperatures are modeled within Williams-Lax (WL) theory\cite{Williams:1951,Lax:1952} using explicit supercells to describe thermal disorder.
Within WL theory, the imaginary part of the temperature- and frequency-dependent dielectric function follows from the statistical average \cite{Williams:1951, Lax:1952, Zacharias:2015},
\begin{equation}
\label{eq:WL-phonon}
\epsilon_{2}(\omega;T) = Z^{-1} \sum_{n} \exp(-\Delta E_{n}\beta)\left \langle \epsilon_{2}(\omega;x) \right \rangle_{n},
\end{equation}
where $Z$=$\sum_{n} \exp(-\Delta E_{n}\beta)$ represents the canonical partition function over $n$ energy states and $\Delta E_{n}$ is the difference in total energy of the disordered structure and the ground state.
$\left \langle \epsilon_{2}(\omega;x) \right \rangle_{n}$ is the quantum mechanical expectation value of the $n$-th snapshot and we represent this by the imaginary part of the dielectric function computed within DFT for that snapshot, see Eq.\ \eqref{eq:optic-inter}.
In this expression, $x$ indicates a set of collective atomic or magnetic disorders, sampled by $n$ different snapshots of the ensemble and $\epsilon_{2}(\omega;x)$ is the dielectric function.

To model atomic displacements $x$ due to finite temperature, two different approaches are used in this work.
First, we adopt the one-shot method by Zacharias and Giustino (ZG), which describes the structural change at finite temperature using a single optimal set of displacements to approximate temperature-induced lattice disorder and also incorporates the zero-point energy motion defined by zero-point vibrational amplitude from the quantum harmonic oscillator solution\cite{Patrick:2014, Zacharias:2016}.
The theoretical details are discussed in Ref.\ \onlinecite{Zacharias:2016}, while the practical implementation of the VASP code is derived from the method elucidated in Ref.\ \onlinecite{Karsai:2018}.
To explore the influence of zero-point vibrations, we compare to a technique that relies on snapshots of displaced atomic geometries which are constructed by superimposing harmonic phonon modes with randomly sampled amplitudes and phases based on classical statistics (CS)\cite{Eckold:2003, Aberg:2013}.
In this method, the mean-square amplitude of mode $i$ is defined as
\begin{equation}
\label{eq:CS-phonon}
\left < \left | Q_j \right |^{2} \right > = \frac{k_{\mathrm{B}}T}{\omega^2_j},
\end{equation}
and the phase is modeled by a cosine function.
The unit of $\left | Q_j \right |^{2}$ is $kg \cdot m^{2}$.
Since in Eq.\ \eqref{eq:CS-phonon} the amplitude is proportional to lattice temperature, vibrational motions of the zero-point energy are not captured in this approach.

To address thermal fluctuations of the magnetic order at finite temperatures we 
sample collective angular disorder $\theta$ applied to the magnetic moments of the material.
This magnetic thermal disorder is described using atomistic spin dynamics based on the stochastic Landau-Lifshitz-Gilbert equation, employing parameters of magnetic interactions derived from DFT,
\begin{equation}
\label{eq:StochasticLLG}
\begin{split}
\frac{d\mathbf{m}_{i}}{dt} = &- \gamma_{L}\mathbf{m}_{i} \times (\mathbf{B}_{i}+\mathbf{B}_{i}^\mathrm{fl})\\ 
&- \gamma_{L} \frac{\alpha}{m_{i}} \mathbf{m}_{i} \times \left[ \mathbf{m}_{i} \times (\mathbf{B}_{i}+\mathbf{B}_{i}^\mathrm{fl}) \right].
\end{split}
\end{equation}
Here $\gamma_{L}$=$\gamma / (1+\alpha^2)$ denotes the renormalized gyromagnetic ratio, $\gamma$ is the gyromagnetic ratio of the electron,
$\alpha$ is a damping parameter, and $\mathbf{m}_{i}$ denotes the magnetic moment at atomic site $i$.
$\mathbf{B}_{i}$ is the effective magnetic field derived from exchange interactions, magnetocrystalline anisotropy, and magnetic dipolar interaction.
The fluctuating magnetic field $\mathbf{B}_{i}^\mathrm{fl}$ introduces the effect of temperature via random fluctuations generated by a Gaussian distribution with an average of zero and variance proportional to the temperature \cite{Eriksson:2017}.
For these simulations, we use the existing implementation in the  \texttt{UppASD}\cite{Eriksson:2017} code.
This classical approach to atomistic spin dynamics does not allow for including the quantum mechanical zero-point magnon energy\cite{Anderson:1952} in our description of the magnetic disorder.

To implement WL theory\cite{Williams:1951, Lax:1952} to compute optical and magneto-optical properties at finite lattice or magnetic temperature, we employ explicit supercell calculations.
For each supercell with temperature-dependent atomic displacements or fluctuating magnetic moments, we compute the complex, frequency-dependent dielectric tensor following Ref.\ \onlinecite{Gajdos:2006},
\begin{equation}
\label{eq:optic-inter}
\begin{split}
\varepsilon_{\alpha \beta}^{(2)}(\omega;x,\theta)=\frac{4 \pi^{2} e^{2}}{\Omega} & \lim_{q\rightarrow 0} \frac{1}{q^{2}} \sum_{c,v,\mathbf{k}} 2 \omega_{\mathbf{k}} \delta (\epsilon_{c\mathbf{k}}^{x,\theta}-\epsilon_{v\mathbf{k}}^{x,\theta}-\omega)\times \\
&\times \left \langle u_{c\mathbf{k}+e_{\alpha}q}^{x,\theta}|u_{v\mathbf{k}}^{x,\theta} \right \rangle \left \langle u_{c\mathbf{k}+e_{\alpha}q}^{x,\theta}|u_{v\mathbf{k}}^{x,\theta} \right \rangle ^{*}.
\end{split}
\end{equation}
Here $\omega$ is the photon frequency, $\alpha$ and $\beta$ are Cartesian axes, $\mathbf{k}$ samples the first Brillouin zone, and $c$ and $v$ are conduction and valence band indices, respectively.
$u_{c\mathbf{k}}^{x,\theta}$ and $u_{v\mathbf{k}}^{x,\theta}$ denote the periodic Kohn-Sham orbitals and $\epsilon_{c\mathbf{k}}^{x, \theta}$ and $\epsilon_{v\mathbf{k}}^{x, \theta}$ represent eigenvalues of these for supercells with atomic ($x$) or magnetic disorder ($\theta$).
To enforce magnetic disorder from atomistic spin dynamics simulations, a constraint of magnetic moments is used that includes a penalty energy term in the total energy, as implemented in \texttt{VASP}\cite{VASPmanual:2022}.
Finite electronic temperature is simulated using occupation number constraints following Eq.\,\eqref{eq:FermiDirac} when evaluating Eq.\ \eqref{eq:optic-inter}.

Therefore, with this approach, it is possible to simultaneously incorporate thermal contributions from electrons, phonons, and magnons by simulating a set of supercell models.
We note that we assume negligible magnon-phonon coupling by imposing magnetic disorder computed at zero lattice temperature also for atomic disorder at finite lattice temperature in the ZG approach.
Furthermore, we adopted frequency-dependent electron lifetimes, quadratically fitted to earlier $GW$ results for Fe\cite{Zhukov:2006} (see Fig.\,S4), to describe electron-electron scattering effects when broadening the optical spectra.
Due to $\mathbf{k}$-point convergence, we apply a lower bound of 0.1\,eV for broadening, see details in section III of the SI.
This frequency-dependent electron lifetime provides a more accurate description of electron-electron scattering lifetimes up to about 6 eV, but it results in the loss of the characteristic peak around 6.7 eV in the measured optical spectra as shown in Fig.\,S5.

From the complex, frequency-dependent dielectric tensor within WL theory, we evaluate optical as well as the first-order magneto-optical coefficient $K$\cite{Hamrlova:2016}
\begin{equation}
\label{eq:first}
    K=\frac{1}{2}(\varepsilon_{\mathrm{xy}}^{[001]}-\varepsilon_{\mathrm{yx}}^{[001]}),
\end{equation}
which we compute using the $xy$ off-diagonal elements of the dielectric tensor since we put the magnetization along the $z$-axis, i.e.\ the [001] direction of the conventional cell.

The optical conductivity spectra can be derived from the dielectric function using\cite{Fox:2010}
\begin{equation}
\label{eq:dielec2op}
    \varepsilon(\omega) = 1 + \frac{i \sigma(\omega)}{\varepsilon_{0}\omega},
\end{equation}
where $\varepsilon_{0}$ represents the vaccum permittivity.
In this paper, we plot all the spectra using $\varepsilon(\omega) \cdot E$, where $E$ is the photon energy, to compare both experimental and computational results with those in Silber's work\cite{Silber:2019}.

\section{\label{sec:comp}Computational Details}

First-principles density functional theory calculations are carried out for structural relaxation of the two-atom ground-state conventional cell and for calculating the electronic and optical properties for all snapshots of disordered supercells at finite temperatures, using the Vienna \emph{Ab-Initio} Simulation Package (\texttt{VASP}) \cite{Kresse:1996,Kresse:1999,Gajdos:2006,Steiner:2016}.
A kinetic energy cutoff of 500\,eV is applied to the plane-wave expansion of the Kohn-Sham states, and the projector-augmented wave method \cite{Blochl:1994} is used to describe the electron-ion interaction.
The exchange-correlation energy term in the DFT Hamiltonian is described using the generalized-gradient approximation parametrized by Perdew, Burke, and Ernzerhof (PBE) \cite{Perdew:1997}.
For structural relaxation, we used a collinear magnetism framework and a $24\times24\times24$ Monkhorst-Pack (MP)  $\mathbf{k}$-point grid \cite{Monkhorst:1976}.

Subsequently, we compute the phonon dispersion (see Fig.\,S1(b) in the supplementary materials) using the finite displacement method implemented in the \texttt{phonopy} package \cite{Togo:2015}.
For phonon dispersion and for the computation of the dielectric tensor in the presence of atomic or magnetic disorder, we use a $3\times3\times3$ supercell with 64 Fe atoms, the noncollinear magnetism framework with spin-orbit coupling \cite{Steiner:2016}, and a randomly shifted $\Gamma$-centered $6\times6\times6$ $\mathbf{k}$-point grid.
The convergence of (magneto-)optical spectra has been checked with respect to supercell size, shown in Fig.\,S13 and $\mathbf{k}$-point sampling, see Fig.\,S14, and we find that the imaginary part of the optical and first-order magneto-optical spectra are converged to within about 10\,\% and 25\,\%, respectively, for photon energies above 0.5\,eV.
Due to the small signal of second-order magneto-optical spectra, achieving converged results is much more challenging (see details in SI).

Exchange interaction coefficients of BCC Fe are calculated using multiple scattering theory through the Korringa-Kohn-Rostoker (KKR) method\cite{Korringa:1947, Kohn:1954}.
A practical implementation in this work is achieved using the spin-polarized relativistic \texttt{SPR-KKR} code \cite{Ebert:2011} to compute isotropic exchange coefficients using the approach developed by Lichtenstein \emph{et al.}\cite{Liechtenstein:1984}
The spin Hamiltonian is
\begin{equation}
\label{eq:exchange}
\mathcal{H}_{ex} = - \sum_{i \neq j}J_{ij}e_{i}e_{j},
\end{equation}
where $i$ and $j$ are atomic site indices, $J_{ij}$ represents the isotropic exchange coefficients, and $e_{i}$ denotes a normalized magnetic moment vector.
In these simulations, the first Brillouin zone is sampled with 2000 random $\mathbf{k}$-points.
We extract all exchange coefficients within the relative distance of $d/a=4$, where $d$ is a spin-spin distance and $a$ is a lattice parameter, see Fig.\,S1(a).
Based on the exchange coefficients from \texttt{SPR-KKR}, atomistic spin dynamics simulations are carried out at finite temperature for a single $9\times9\times9$ supercell using the stochastic Landau-Lifshitz-Gilbert equation implemented in \texttt{UppASD}\cite{Eriksson:2017}.
This cell size resulted from convergence tests of sublattice magnetization and heat capacity, see Fig.\,S2.

For the computation of the dielectric tensor within DFT, we divided the magnetic disorder in the $9\times9\times9$ supercell used for atomistic spin dynamics simulations into a total of 27 different $3\times3\times3$ supercells.
For atomic displacements, we used a single snapshot of a $3\times3\times3$ supercell from the ZG method and employed 27 different $3\times3\times3$ supercells generated by the CS method.
To verify the reproduction of atomic and magnetic disorder distributions using $3\times3\times3$ supercells, we compare the pair (angle) distribution functions with $6\times6\times6$ and $9\times9\times9$ supercells, as illustrated in Fig.\ S11 of the SI.
We note that converging long-wavelength lattice or spin waves requires large supercell sizes that are not achievable in our calculations.
This limits the accuracy of our results at low photon energies (see details in Fig.\ S13 of the SI).
We average the supercell optical spectra based on WL theory\cite{Williams:1951, Lax:1952} and also extract standard errors over the snapshots\cite{Knoop:2023} to estimate the error bar as shown in Fig.\,S12.

\section{\label{sec:results}Results}

\begin{figure}
\includegraphics[width=0.98\columnwidth]{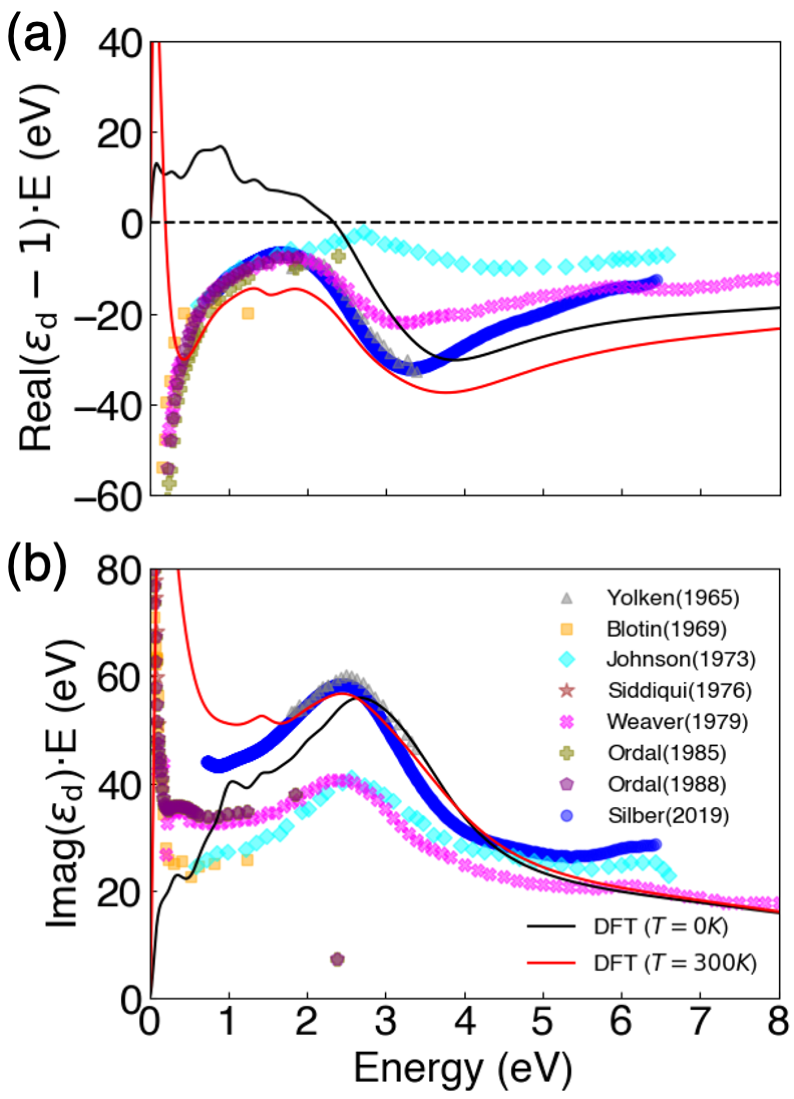}
\caption{\label{fig:diag}(Color online.)
(a) Real and (b) imaginary parts of the averaged diagonal components of the optical spectra, $(\varepsilon_{\mathrm{d}}-1) \cdot E$, of ferromagnetic BCC Fe.
DFT results for 0\,K (without zero-point displacement, black solid line), and 300\,K (electron, lattice, and magnetic temperature, red solid line).
Markers display measured values at room temperature by Yolken \emph{et al.}\cite{Yolken:1965}, Blotin \emph{et al.}\cite{Blotin:1969}, Johnson \emph{et al.}\cite{Johnson:1974}, Siddiqui \emph{et al.}\cite{Siddiqui:1977}, Ordal \emph{et al.}\cite{Ordal:1985, Ordal:1988}, Silber \emph{et al.}\cite{Silber:2019}, and 4.2\,K by Weaver \emph{et al.}\cite{Weaver:1979}
}
\end{figure}

Optical spectra of ferromagnetic BCC Fe computed at $T$=0\,K and at $T$=300\,K of electron, lattice, and magnetic temperatures are plotted together with experimental results \cite{Yolken:1965, Blotin:1969, Johnson:1974, Siddiqui:1977, Ordal:1985, Ordal:1988, Silber:2019, Weaver:1979} in Fig.\,\ref{fig:diag}.
For comparison with polycrystalline experimental results\cite{Yolken:1965, Blotin:1969, Johnson:1974, Siddiqui:1977, Ordal:1985, Ordal:1988, Silber:2019, Weaver:1979}, the diagonal components of the dielectric tensor are averaged when computing the dielectric functions.
The $T$=0\,K result is computed without zero-point displacements, corresponding to the standard approach for computing optical spectra used in the literature.
The real part of the optical spectra is connected to the imaginary part via the Kramers-Kronig transform and, hence, exhibits similar temperature dependence (see also Fig.\,S6).
The maximum standard deviation of the imaginary part over the snapshots at $T$=300\,K is about 0.7\,\%, as shown in Fig.\,S12(a) and (b).
We note that the magnitude of the different measured imaginary parts differ while the real parts show good agreement with each other, which might originate from different surface preparation techniques, as elucidated by Yolken \emph{et al.}\cite{Yolken:1965}
Overall, our results for the optical spectra at $T$=300\,K show good agreement with measured spectra.

More specifically, from Fig.\ \ref{fig:diag} we identify three characteristic changes in optical spectra caused by finite temperature.
First, the positive real part of the optical spectra that we compute for $T$=0 K at energies less than 3\,eV becomes negative after introducing the finite temperature in the simulations.
This causes a broad peak to form between 0.5 and 2.5 eV bringing the real part in good agreement with experiment both in terms of peak position and magnitude.
Second, the simulation result for $T$=300 K shows a very large peak of the imaginary part of the optical spectra below 1\,eV that is absent in the $T$=0 K simulation.
We attribute this to phonon and magnon-assisted intraband transitions that can be captured by the changes of optical matrix elements in supercell simulations with thermal disorder, as analyzed in detail in Sec.\ \ref{sec:disscusion}.
Such intraband response is also reported for measured spectra\cite{Blotin:1969, Siddiqui:1977, Ordal:1985, Ordal:1988, Weaver:1979}, however, in a smaller range of low energies.
The diverging signal of the real part below 0.5\,eV and the exaggerated width of the intraband signal in the imaginary part may in part be due to remaining finite-size effects in our simulations, rendering a description of long-wavelength lattice vibrations and magnetic fluctuations challenging.
Third, we identify a redshift of about 0.2 eV of the peak near 2.7\,eV in the imaginary part of the optical spectra at 0\,K, caused by the shifting of the electronic band structure with thermal demagnetization.
The physical origin of all three effects is discussed in detail in Sec.\,\ref{sec:disscusion}.

\begin{figure*}
\includegraphics[width=0.98\textwidth]{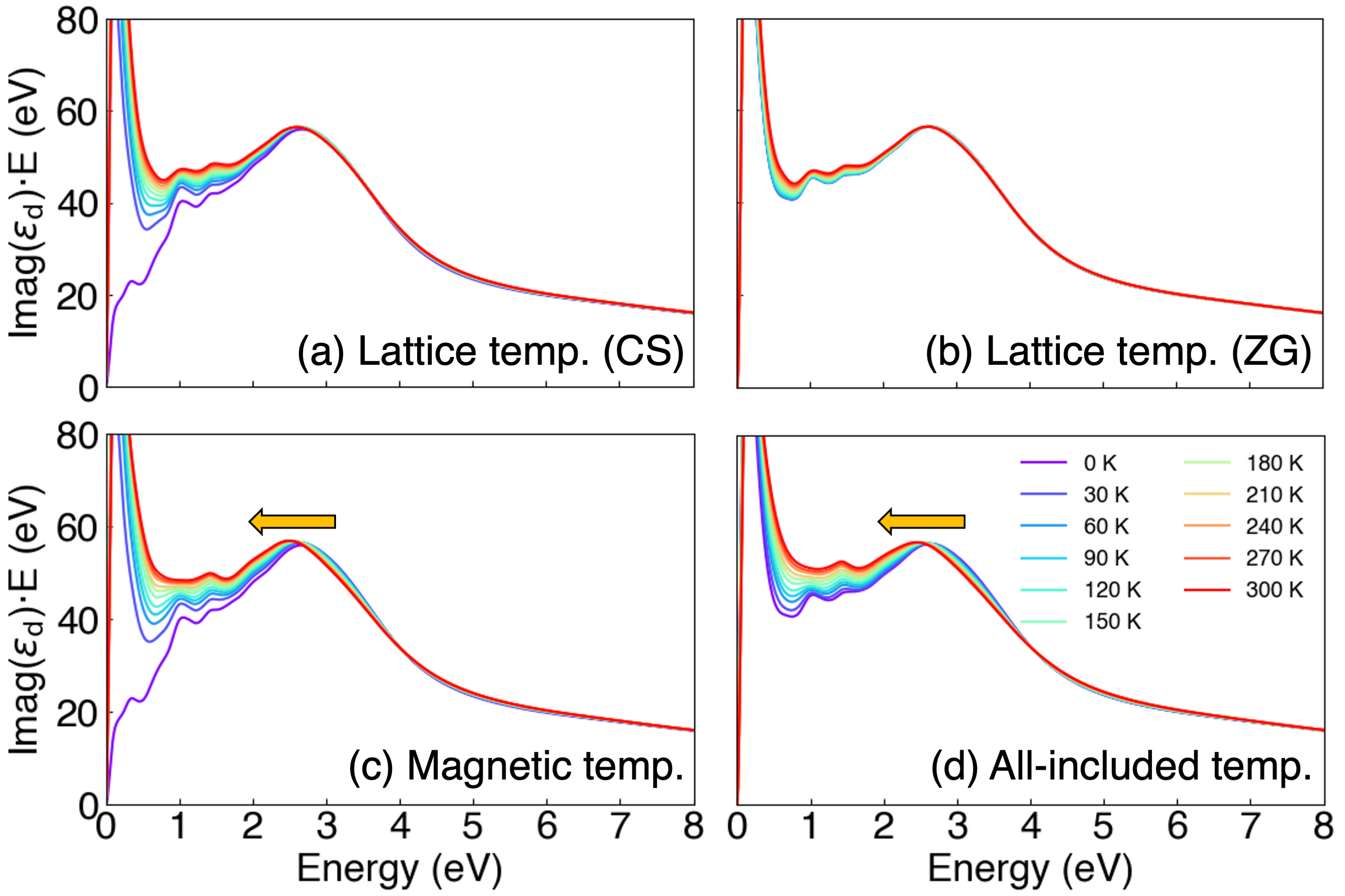}
\caption{\label{fig:imag-temp}(Color online.)
The imaginary part of the averaged diagonal components of the temperature-dependent optical spectra, $(\varepsilon_{\mathrm{d}}-1) \cdot E$, of ferromagnetic BCC Fe from DFT with (a) lattice temperature from classical statistics (CS), (b) lattice temperature from the one-shot Zacharias and Giustino (ZG) method, (c) magnetic temperature with the ground-state atomic structure, and (d) electron, lattice, and magnetic temperature from 0\,K to 300\,K.
Yellow arrows in (c) and (d) indicate a redshift of the peak near 2.7\,eV.
}
\end{figure*}

To analyze these individual contributions due to electron, lattice, and magnetic temperature, we implement them separately when computing optical spectra.
First, Fig.\,S3 in the supplementary material  demonstrates the subtle relationship between electronic temperature and optical spectra, highlighting a very weak effect.
The maximum change is less than 0.2\,\% because an electron temperature of 300\,K only slightly modifies electronic occupations near the Fermi level.
However, the averaged diagonal components of the temperature-dependent optical spectra in Fig.\,\ref{fig:imag-temp} show that the impact of lattice and magnetic temperature on the optical spectra is significant.

Comparing the two different approaches to model lattice temperature shown in Fig.\,\ref{fig:imag-temp}(a) and (b) delineates the effect of zero-point vibrational motions.
Without zero-point motion, there is no signal from intraband transitions at 0\,K in Fig.\,\ref{fig:imag-temp}(a), but these contributions become significant as soon as the lattice  temperature is non-zero.
Including zero-point motion via the ZG method, shows a large peak at low energies independent of temperature in Fig.\,\ref{fig:imag-temp}(b) due to intraband transitions.
As the temperature approaches 300\,K, this quantum mechanical feature becomes less important and the data in Fig.\ \ref{fig:imag-temp}(b) looks very similar to the harmonic sampling results based on classical statistics in Fig.\ \ref{fig:imag-temp}(a).

Figure\ \ref{fig:imag-temp}(c) illustrates the effect purely contributed by magnetic temperature.
Again, at 0\,K no signal from intraband transitions is observed due to the absence of zero-point energy motion in spin waves in this work.
Our results how a large contribution of magnon-assisted intraband transitions immediately as the magnetic temperature increases above 0 K, similar to the lattice temperature from classical statistics in Fig.\ \ref{fig:imag-temp}(a).
Furthermore, we show that a red shift of the peak near 2.7\,eV occurs only when introducing magnetic temperature.
Such a feature does not occur in Fig.\ \ref{fig:imag-temp}(a) which shows that it is independent of lattice temperature.
Our analysis thus shows that the large optical spectra below 1.0\,eV originates from quantum-mechanical zero-point vibrations, while the redshift of the peak near 2.7\,eV is dominated by magnetic temperature.
In this work, we only capture zero-point lattice vibrations, but the magnetic system can have a similar zero-point effect and both together dominate the temperature dependence below 1.0\,eV.
Since Fig.\ \ref{fig:imag-temp} shows a small difference between the CS and ZG results at energies above 2 eV and because the magnetic peak redshift is much higher than the lattice peak shift (see explicit discussion of Fig.\ \ref{fig:peak} below) we conclude that if zero-point magnon effects were included, that redshift persists.
We also expect a smaller spread of the curves between 0.5 eV and 2 eV from the comparison of CS and ZG results in that energy range.

\begin{figure}
\includegraphics[width=0.98\columnwidth]{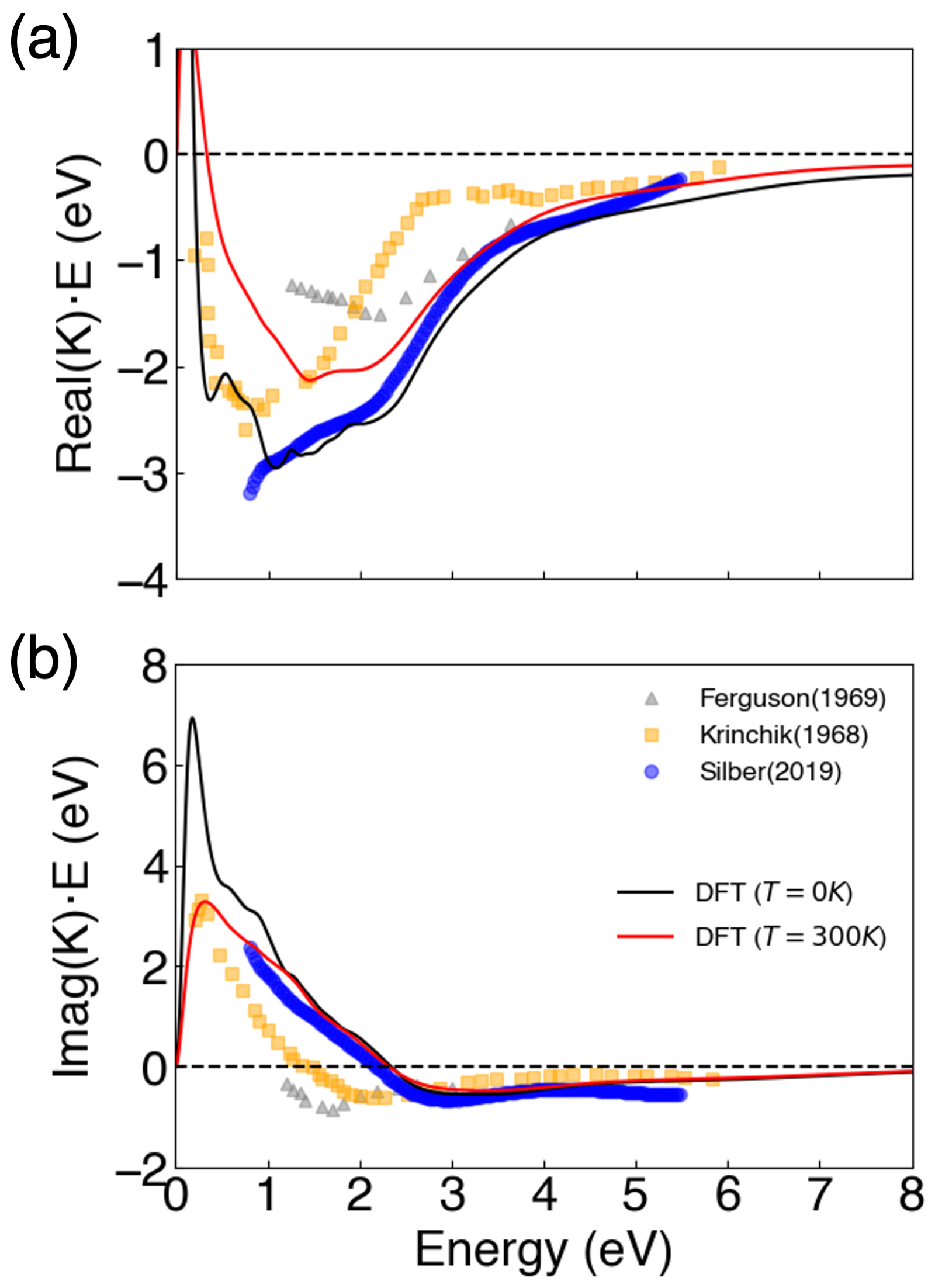}
\caption{\label{fig:first}(Color online.)
(a) Real and (b) imaginary parts of the first-order magneto-optical spectra, $K \cdot E$, of ferromagnetic BCC Fe from DFT results with 0\,K (black solid line, including zero-point motion via the ZG approach) and electron, lattice (ZG), and magnetic temperature of 300\,K (red solid line). 
Markers display measured values at room temperature by Ferguson \emph{et al.}\cite{Ferguson:1969}, Krinchik \emph{et al.}\cite{Krinchik:1968}, and Silber \emph{et al.}\cite{Silber:2019}
}
\end{figure}

Next, we computed real and imaginary parts of the first-order magneto-optical spectra of ferromagnetic BCC Fe (see Fig.\,\ref{fig:first}).
Our spectra display typical demagnetization behavior of reduced magneto-optical signals over the photon energy range as the temperature increases, corresponding to a 10\,\% reduction of the net magnetization in DFT (see temperature dependent magnetization data in Fig.\,S2 of the SI ).
Comparison with experimental results shows good agreement.
For the real part in Fig.\,\ref{fig:first}(a), the overall spectral shape matches well, with a slight difference in the peak position around 1.5 eV.
This difference between DFT data at 0\,K and experiments has also been reported in the literature \cite{Oppeneer:1992, Miyazawa:1999, Stejskal:2018, Silber:2019, Stejskal:2021}.
The sharp positive peak of the real part at photon energies below 0.2\,eV might be artificial due to remaining finite size effects for the supercells used in this work.
These arise because the Drude model relies on intraband transitions, that are only captured by band folding in this work.
Thus, the minimum interband transition energy captured in our approach depends on the supercell size and we do not attempt to converge spectra at energies less than about 0.5 eV, corresponding to the intraband-induced increase of the imaginary part of the dielectric function at low energies (see Fig.\ S12 in the SI).

The imaginary part of the first-order magneto-optical spectra in Fig.\ \ref{fig:first}(b) also shows good agreement with measured spectra, including the sign change near 2.5\,eV.
The magnitude of our computed spectrum for $T$=300\,K is smaller than for $T$=0\,K because of thermal demagnetization.
Since the Curie temperature of BCC Fe is around 1043\,K\cite{Ashcroft:1976}, thermal demagnetization at 300\,K is not drastic.
However, our simulations describe a gradual reduction of first-order magneto-optical spectra over the whole photon energy range as temperature increases (see Fig.\,S7 of the SI), roughly corresponding to the theoretical expectation of proportionality of the linear magneto-optical signal and net magnetization.
The standard deviation of first-order magneto-optical spectra due to averaging of magnetic disorder is more discernible at photon energies below 1\,eV than what we found for optical spectra (see Figs.\,S12(c) and (d)), since the first-order magneto-optical spectra signal is about 10 times smaller.
While this error does not affect our discussed conclusions, it requires careful convergence with respect to the number of snapshots used.

We also attempted to calculate second-order magneto-optical spectra, $G_{s} \cdot E$, see Fig.\,S8, but they suffer from a large standard deviation over the entire photon energy range, as plotted in Fig.\,S12(e) and (f).
The reason for this is that second-order magneto-optical signals are about 100 times smaller than optical spectra, $(\varepsilon_{\mathrm{d}}-1) \cdot E$, requiring much higher accuracy when averaging magnetic disorder.
In addition, these calculations need to be implemented using very dense $\mathbf{k}$-point grids as shown in Fig.\,S14(e) and (f).
This is also discussed in Ref.\,\onlinecite{Stejskal:2018} and the authors show that sufficiently dense $\mathbf{k}$-point grid can provide appropriately converged optical spectra at 0\,K.

\section{\label{sec:disscusion}Discussion}

\begin{figure}
\includegraphics[width=0.98\columnwidth]{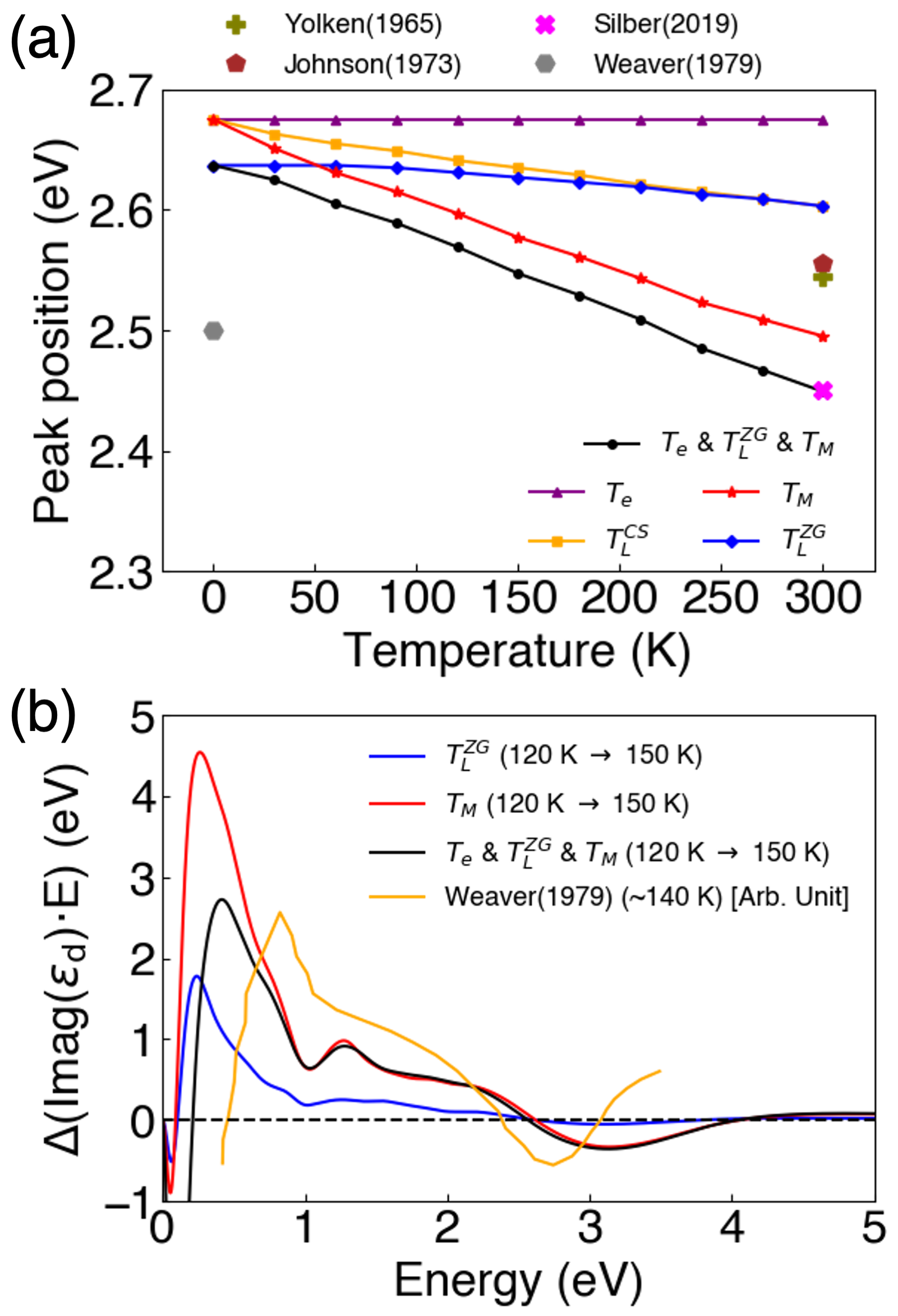}
\caption{\label{fig:peak}(Color online.)
(a) Position of the peak in the imaginary part of the optical spectrum, $(\varepsilon_{\mathrm{d}}-1) \cdot E$, near 2.7\,eV as a function of temperature.
Electron temperatures (purple solid line, $T_{e}$), classical-statistics lattice temperatures (orange solid line, $T_{L}^{CS}$), lattice temperatures with zero-point vibrations (blue solid line, $T_{L}^{ZG}$), magnetic temperatures (red solid line, $T_{M}$), and all temperatures combined (black solid line, $T_{e}$ \& $T_{L}^{ZG}$ \& $T_{M}$) are plotted separately.
Measured values at room temperature by Yolken \emph{et al.}\cite{Yolken:1965}, Johnson \emph{et al.}\cite{Johnson:1974}, Weaver \emph{et al.}\cite{Weaver:1979}, and Silber \emph{et al.}\cite{Silber:2019} are plotted with different markers.
(b) Change of the optical spectra, $(\varepsilon_{\mathrm{d}}-1) \cdot E$, due to excursions of lattice and magnetic temperatures from 120 K to 150 K.
Weaver measured the spectral change of the optical spectrum at 140\,K with an arbitrary temperature excursion\cite{Weaver:1979}. 
Its magnitude is in arbitrary units and is rescaled for comparison.
}
\end{figure}

In the following, we analyze the lattice and magnetic contribution to the temperature dependence of the optical-conductivity peak near 2.7\,eV in more detail.
The electronic temperature is not discussed due to its negligible effect on the spectrum (see Fig.\,S3 in the supplementary materials).
Figure \ref{fig:peak}(a) shows that the zero-point lattice vibrations (included in blue and black solid lines) induce a redshift of this peak by about 0.04 eV.
The data also shows that the effect of zero-point vibrations vanishes around 240\,K, which means that the quantum mechanical nuclear motion becomes analogous to classical atomic motion, evident from the converging yellow and blue solid lines.
An analogous behavior was reported for the temperature-dependent band gap of silicon, where the difference also vanishes at high temperatures\cite{Zacharias:2020}.
In the classical approach that neglects zero-point vibrations (yellow solid line), the red shift is about 0.07\,eV.
Hence, we conclude that the effect due to lattice temperature is very weak.
Nevertheless, the curve that includes all temperature effects (black circles) differs from that for the magnetic temperature (red stars) by about 0.04\,eV across the entire temperature range.
Figure \ref{fig:peak}(a) strikingly shows that while the absolute peak position depends on the zero-point lattice vibrations, its near-linear temperature dependence is dominated by the magnetic temperature.
This dependence of the peak position on magnetic temperature can serve as an experimental measure of the degree of magnetic disorder in ferromagnetic BCC Fe.

Importantly, the discrepancy in position between the calculated peak at 0\,K, excluding zero-point vibrational motions, and the measured peak at 300\,K is solved by the introduction of temperature.
In Fig.\,\ref{fig:peak} (a), the calculated peak position at 300\,K (black solid line) is now placed in the range of measured values at 300\,K\cite{Yolken:1965, Johnson:1974, Silber:2019}.
However, these measured peak positions at 300\,K\cite{Yolken:1965, Johnson:1974, Silber:2019} vary from 2.45 eV to 2.56 eV and the peak position at 4.2\,K measured by Weaver \emph{et al.}\cite{Weaver:1979} is also around 2.5\,eV, making it difficult to diretly conclude the redshifting behavior from experiment.

However, comparison with the measured \emph{change} of the optical spectrum\cite{Weaver:1979} supports the existence of this redshift:
Weaver \emph{et al.}\ performed measurements using a temperature excursion at 140\,K caused by a thin film heater as shown in Fig.\,\ref{fig:peak}(b).
Their data provides evidence of a peak redshift via a sign flip near 2.4\,eV\cite{Weaver:1979}, implying that the signal at the original peak position decreases while the signal at the new peak position increases as the peak moves left.
Our calculated change of the optical spectrum for magnetic temperatures of 120\,K and 150\,K, similar to the temperature of 140 K used in experiment \cite{Weaver:1979}, 
shows a similar sign change near 2.6\,eV, in qualitatively good agreement with Weaver's result, including the height ratio of the peaks observed at about 0.9\,eV and 2.8,eV, see Fig.\,\ref{fig:peak}(b).
Notably, the result with varying lattice temperature alone does not exhibit such a feature.
We attribute differences between theory and experiment in Fig.\ \ref{fig:peak}(b) at energies below 1 eV to finite-size effects in our simulations.
The slight disagreement of the position of the negative peak near 2.8 eV might be explained by the use of a local density approximation in our simulations and better descriptions of the electron-electron interaction might improve the agreement, based on the reduction of bandwidth in the previous $GW$ studies\cite{Muller:2019, Nabok:2021}.

\begin{figure*}
\includegraphics[width=0.98\textwidth]{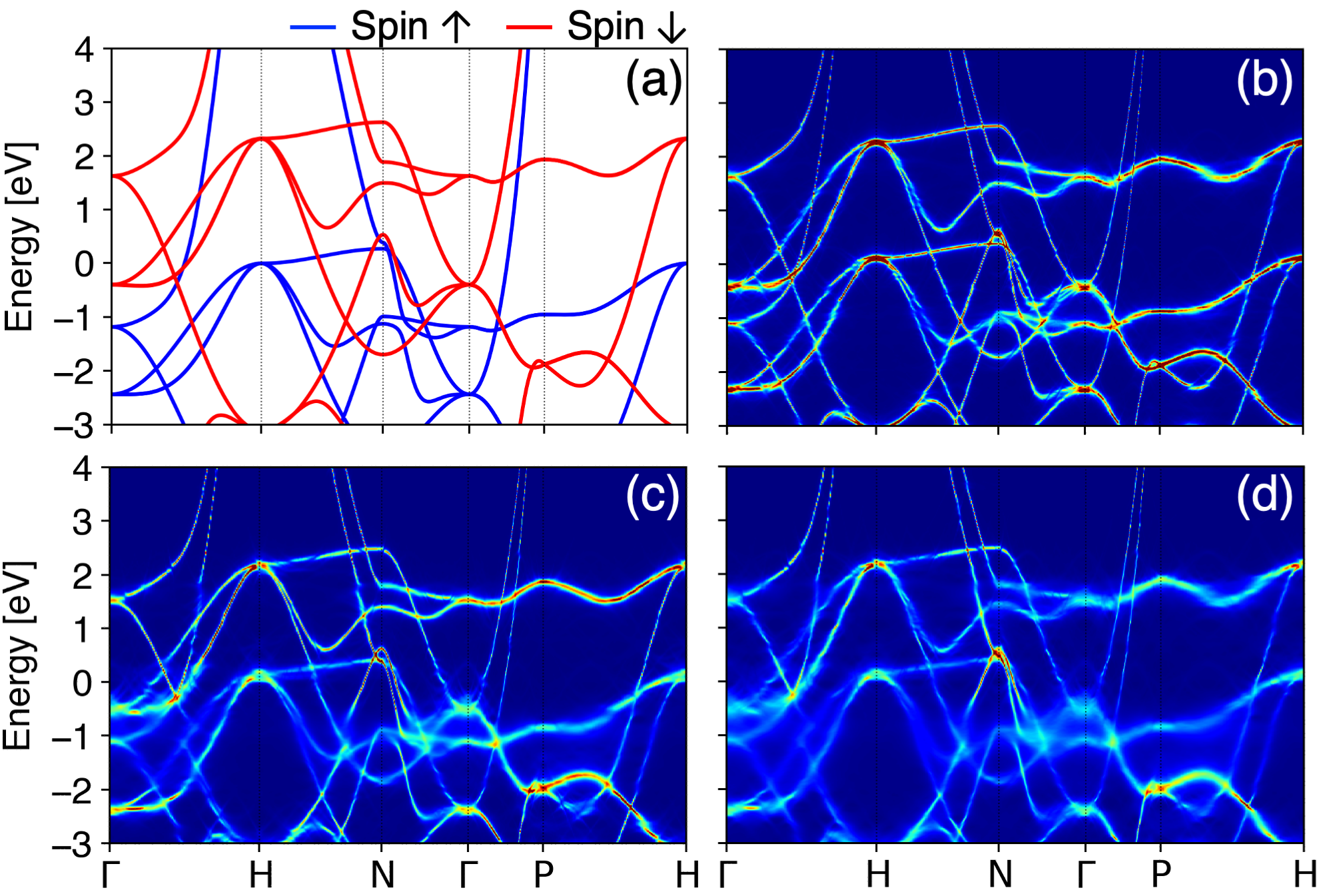}
\caption{\label{fig:band}(Color online.)
(a) Spin-polarized electronic band structure of a primitive cell of ferromagnetic BCC Fe, without spin-orbit coupling.
Majority and minority spin states are plotted with blue and red solid lines, respectively.
Unfolded electronic band structure of ferromagnetic BCC Fe averaged over supercell configurations at (b) lattice (ZG) temperature, (c) magnetic temperature, and (d) electron, lattice (ZG), and magnetic temperature of 300\,K.
The Fermi level is set to zero.
}
\end{figure*}

The temperature-dependent electronic band structure of ferromagnetic BCC Fe can elucidate the exact physical origin of the peak red shift.
We analyze characteristic band anomalies, such as kinks, caused by electron-phonon and electron-magnon scattering.
Additionally, we can identify the reduction of exchange splitting as the main cause.
Due to band folding, electronic band structures of supercells are too complicated to interpret as temperature changes.
Thus, we adopted the effective band structure method developed by Popescu \emph{et al.}\cite{Popescu:2012} to extract the unfolded band structure in terms of the reciprocal space of the primitive unit cell.
Our data in Fig.\ \ref{fig:band}(a) plots the spin-polarized band structure at 0\,K, indicating a clear exchange splitting due to the ferromagnetic ordering.
This agrees well with the unfolded band structure at an electronic temperature of 300\,K displayed in Fig.\,S9(a), due to the small effect of electronic temperature.
Both lattice and magnetic temperature of $T$=300 K introduce a renormalization of the electronic band structure due to thermal disorder, as can be seen from renormalized bandwidths in Fig.\ \ref{fig:band}(b)-(d).
This introduces the broadening of optical spectra due to electron-phonon and electron-magnon scattering.

Figure \ref{fig:band}(b) shows that the renormalization due to the lattice temperatures is not significant and the unfolded band structure generally maintains its shape and gaps between energy states.
As a result, lattice temperature leads to no drastic change besides the appearance of intraband transitions in the optical spectra, consistent with our discussion above.
Conversely, Fig.\,\ref{fig:band}(c) shows that magnetic temperature induces two unique temperature-dependent behaviors.
First, both spins shift oppositely, in that the majority spin states corresponding to blue solid lines in \ref{fig:band}(a) increase their energy while minority spin states corresponding to red solid lines in \ref{fig:band}(a) decrease it.
This trend can be clearly seen in the animation that is part of the supplementary materials.
This behavior of the spin-polarized electronic band structure indicates a reduction of the exchange splitting due to thermal demagnetization, which is comparable to a red shift of the peak position in the imaginary part of the optical spectrum.
Therefore, we conclude that the red shift of the peak near 2.7 eV originates from the reduction of exchange splitting in the electronic band structure due to thermal demagnetization.

\begin{figure}
\includegraphics[width=0.98\columnwidth]{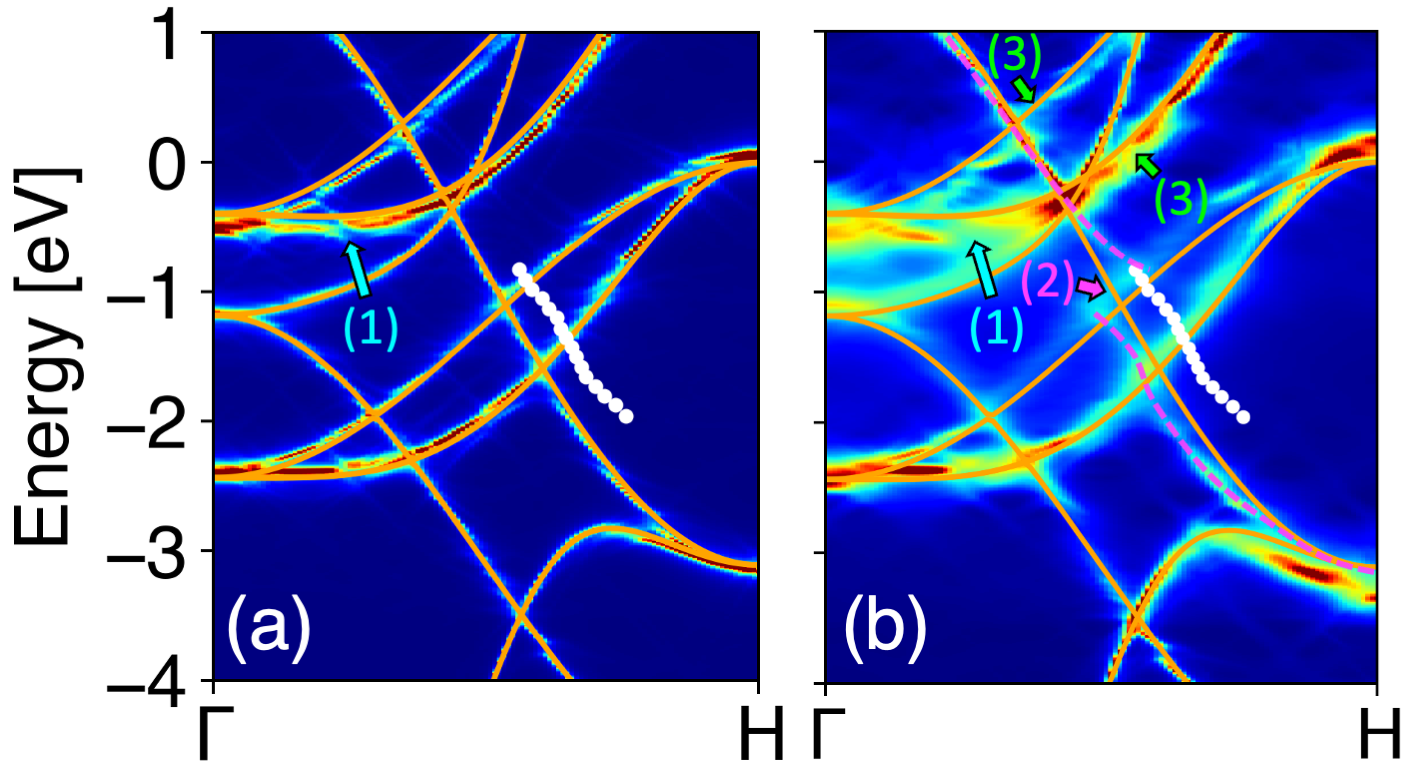}
\caption{\label{fig:band-zoom}(Color online.)
Unfolded electronic band structure between $\Gamma$ and H of ferromagnetic BCC Fe at (a) lattice (ZG) temperature of 300\,K and (b) electron, lattice (ZG), and magnetic temperature of 300\,K.
Orange solid lines represent the electronic band structure of a primitive cell without the spin-orbit coupling effect.
Anomalous band kinks are labeled with (1), (2), and (3).
The magenta dashed line highlights kink (2) in the unfolded electronic band structure.
White circles show the results from the photoemission experiment in Ref.\,\onlinecite{Mylnczak:2019}, which matches with the lower part of the magenta dashed line.
This figure magnifies the data in Fig.\ \ref{fig:band} and enhances the color scale.
}
\end{figure}

Next, we analyze the band structure between $\Gamma$ and $H$ in more detail in Fig.\ \ref{fig:band-zoom}.
First, we observe strong renormalization due to lattice temperature and magnetic temperature in the energy range from $-0.5$\,eV to $-2.0$\,eV, see Fig.\,\ref{fig:band-zoom} as can be seen from the smeared out bands.
We also analyze the appearance of characteristic kinks within 2.0\,eV of the Fermi level as pointed by arrows (1), (2), and (3) in Fig.\,\ref{fig:band-zoom}.
These kinks are absent at 0\,K in Fig.\,\ref{fig:band} (a) or at finite electronic temperature (see Fig.\,S9(a)), but occur as unique features associated with both lattice and magnetic temperature.
There is a weak feature, labeled (1) in Fig.\ \ref{fig:band-zoom}, that can be attributed to finite lattice temperature, and stronger ones, labeled (2) and (3), that appears for finite magnetic temperature.
The size of these second features near $-1.0$\,eV between $\Gamma$ and $H$ proportionally increases as magnetic temperature rises, see the animated figure file in the supplementary materials.
The magnitude of the band kink in our work might be overestimated because the predicted Curie temperature of 880\,K from atomistic spin dynamics (see Fig.\,S2) is underestimated compared to the measured value of 1043\,K\cite{Ashcroft:1976}, inducing a somewhat larger impact of electron-magnon scattering at room temperature in our simulations.

Previous investigations\cite{Muller:2019, Mylnczak:2019, Nabok:2021} reported some band anomalies of BCC Fe that we also observed.
Studies using the $GW$+$T$ matrix many-body Green's function method\cite{Muller:2019, Nabok:2021} presented the band renormalization and band anomalies after introducing electron-magnon coupling.
For example, band kinks, labeled (3) in Fig.\,\ref{fig:band-zoom}, were also observed in their simulations, while they didn't encounter feature (1) since their study didn't introduce any lattice temperature effects\cite{Muller:2019, Nabok:2021}.
This comparison demonstrates that our method can capture the electron-magnon coupling effects by introducing magnetic disorder.
The band kinks shown in our result but not in the $GW$+$T$ methods imply that our method also can capture the electron-phonon coupling effects after introducing atomic disorder.
Furthermore, the $GW$+$T$ method encompasses electron-electron coupling effects that reduce the overall bandwidth\cite{Muller:2019, Nabok:2021}, a consideration not accounted for in our method.
For feature (2), there is no band kink but a \emph{waterfall} structure in their study\cite{Muller:2019, Nabok:2021}, which is also measured in photoemission experiments\cite{Mylnczak:2019} as shown as white circle markers in Fig.\,\ref{fig:band-zoom}.
This \emph{waterfall} structure shows a qualitatively good agreement with the lower part of the magenta dashed line of feature (2) in Fig.\,\ref{fig:band-zoom} from our result.
Finally, our method includes spin-orbit coupling effects, that are neglected in the $GW$+$T$ study.
However the lack of band kink (2) in the $GW$+$T$ results does not originate from the lack of spin-orbit coupling since it still appears when we exclude the spin-orbit coupling effect, which requires further investigation.

\begin{figure}
\includegraphics[width=0.98\columnwidth]{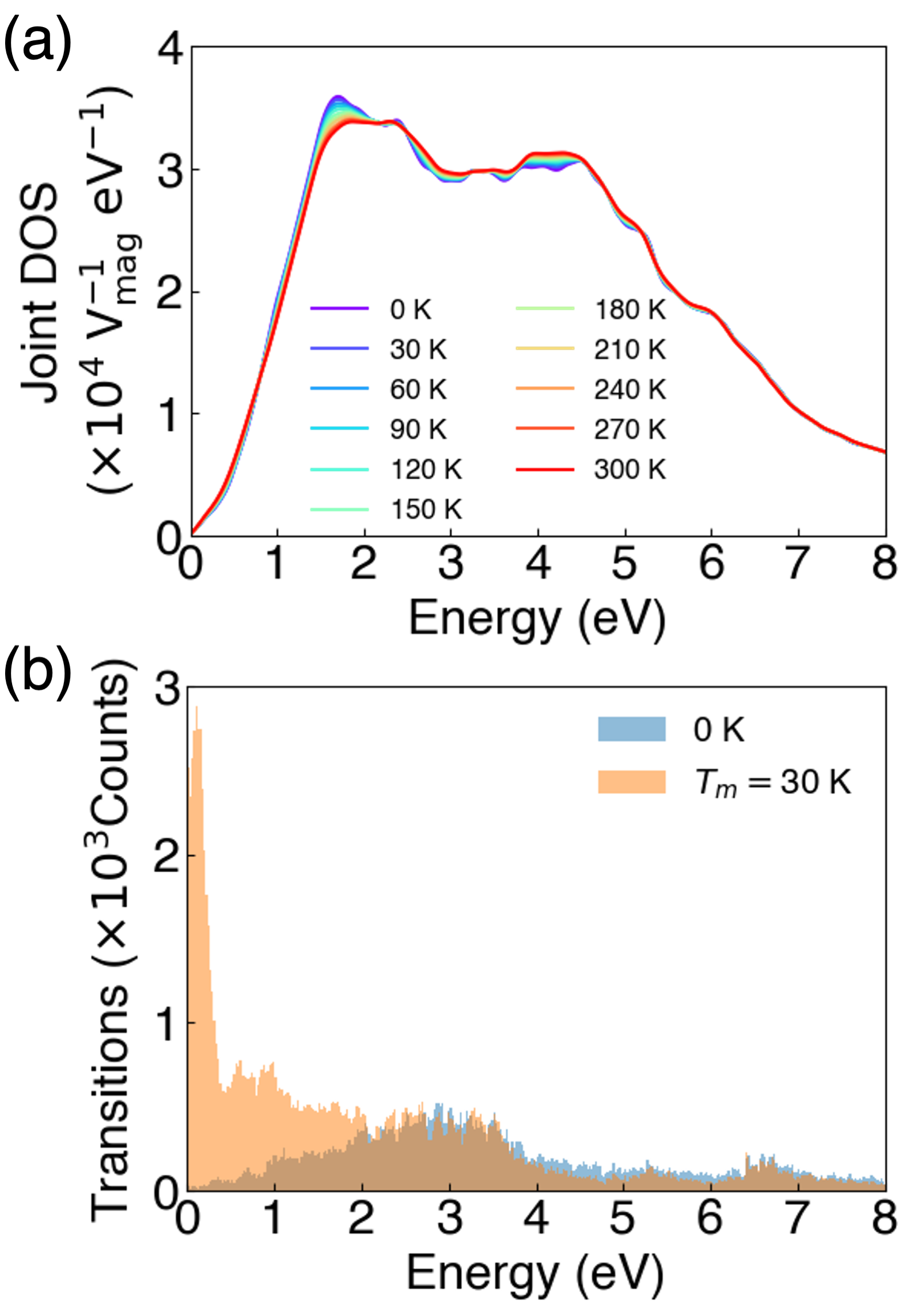}
\caption{\label{fig:factor}(Color online.)
(a) Joint density of states for different magnetic temperatures.
(b) Population distribution of transitions with dipole matrix elements $p_{x}^{2}>0.05$ vs.\ transition energy between conduction and valence bands at 0\,K (without zero-point displacement, cyan bars) and a magnetic temperature of 30\,K (orange bars).
}
\end{figure}

We also analyze the origin of phonon and magnon-assisted intraband transitions at energies below 2\,eV in more detail.
Unlike the perturbative AH\cite{Allen:1976} and HBB theories\cite{Hall:1954} discussed in the introduction, the supercell approach to including thermal disorder accounts for both temperature-induced modifications of electronic energies and Kohn-Sham states beyond the perturbative regime.
Hence, it also accounts for changes in the optical transition matrix elements in the computation of the optical spectrum through Eq.\,\eqref{eq:optic-inter}.
Temperature-induced changes of the electronic energies and optical matrix elements are disentangled using the joint density of states (JDOS) and the population distribution illustrated in Fig.\ \ref{fig:factor}, respectively.

The JDOS in Fig.\,\ref{fig:factor}(a) shows only a minute change as a function of magnetic temperature and a similarly subtle change is also observed for finite lattice temperature.
This implies that the change of the optical spectrum is mediated by large modifications of optical dipole matrix elements.
When we plot the entire population distribution of the optical transition matrix elements as shown in Fig.\,S10, there is no apparent difference between 0\,K and magnetic temperature of 30\,K, showing that their total number and energy distribution does not change.
Conversely, Fig.\,\ref{fig:factor}(b) illustrates that \emph{large} optical dipole matrix elements (e.g.\  $p^{2}_{x} > 0.05$) occur much more often below transition energies of up to 2\,eV for non-zero magnetic temperature.
This shows that the corresponding intraband transitions at zero temperature are dipole forbidden, but become dipole allowed due to magnetic or lattice excitations in the presence of disorder.
Because this signal does not appear without any disorder, we interpret them as intraband transitions.
Such optical transitions with momentum transfer due to band folding, require a sufficiently large supercell to sample long-wavelength phonon and magnon modes related to the transitions with small photon energy near the Fermi level to prevent the finite size effect.

\begin{figure}
\includegraphics[width=0.98\columnwidth]{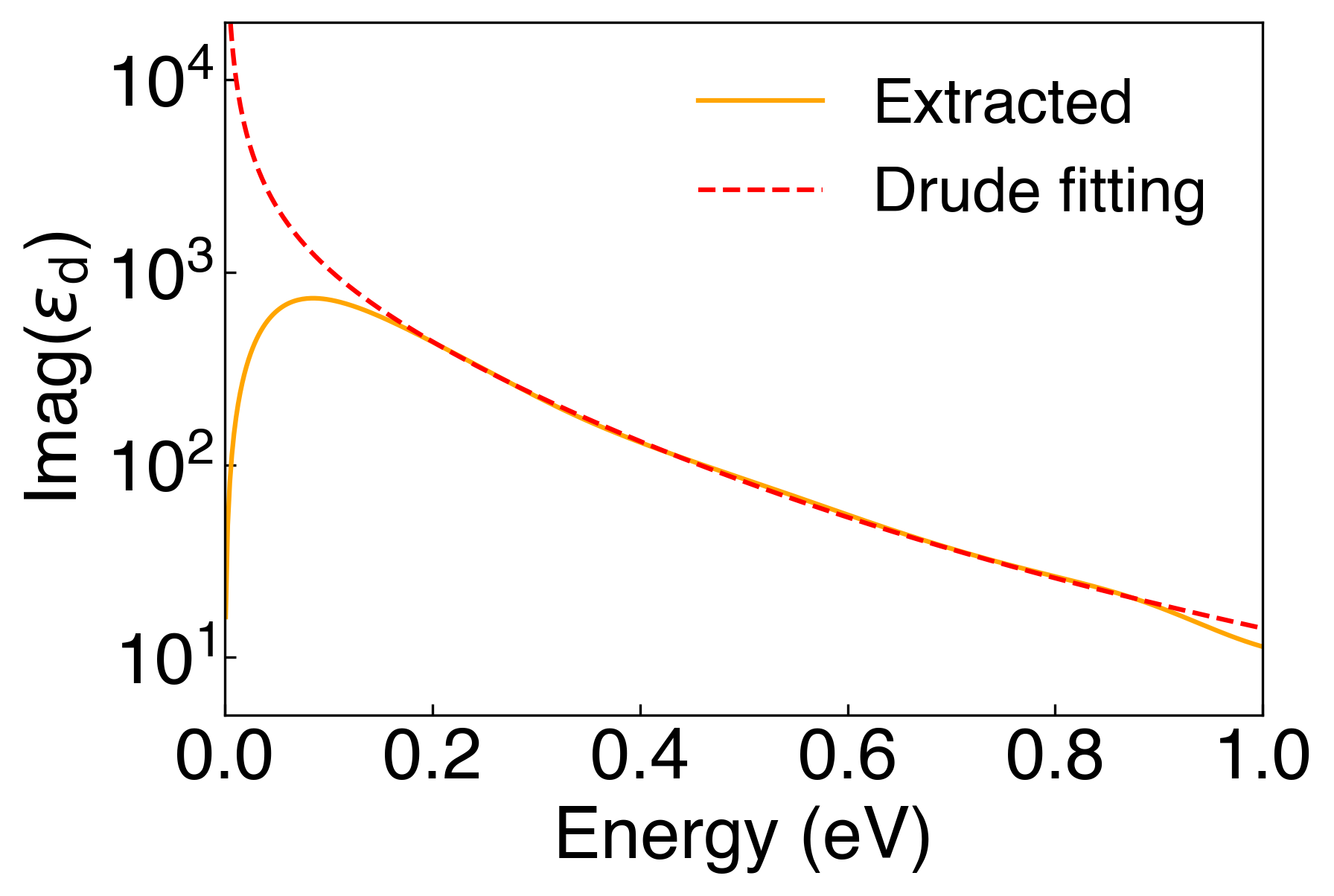}
\caption{\label{fig:drude}(Color online.)
The intraband transition term (orange solid line) of the imaginary part of the calculated dielectric function is extracted by subtracting the dielectric function at lattice (CS) temperature of 0\,K from that at electron, lattice (ZG), and magnetic temperatures of 300\,K. The Drude fitting result is shown as a red dashed line. The fitting is implemented using the extracted intraband transition term from 0.2\,eV to 0.9\,eV.
}
\end{figure}

Finally, we return to the question of modeling indirect transitions that are expected to be particularly important in indirect semiconductors\cite{Zacharias:2015, Zacharias:2016}, direct forbidden semiconductors, and metals.
As discussed in the previous paragraph, a disorder due to temperature or zero-point vibrations causes symmetry breaking that enables otherwise dipole-forbidden optical transitions that are not accounted for in simulations using primitive cells.
It was discussed before that this approach is capable of capturing optical transitions in indirect semiconductors\cite{Zacharias:2015, Zacharias:2016}.
To explore whether it also captures Drude-like intraband transitions in a metallic system, we compare it to the Drude model \cite{Ordal:1985}
\begin{equation}
\label{eq:drude}
\mathrm{Im}(\varepsilon_{\mathrm{d}}) = \frac{\omega_{\mathrm{\tau}}\omega_{\mathrm{p}}^{2}}{\omega(\omega^{2}+\omega_{\mathrm{\tau}}^{2})},
\end{equation}
where $\omega$, $\omega_{\mathrm{p}}$, and $\omega_{\mathrm{\tau}}$ stand for the photon frequency, the plasma frequency, and the damping frequency, respectively.
Previously, Drude parameters $\omega_{\mathrm{p}}$ and $\omega_{\mathrm{\tau}}$ have been extracted by fitting this equation to the low photon frequency range of the imaginary part of the dielectric function in experiments\cite{Ordal:1985, Mendelsberg:2012} and simulations\cite{Desjarlais:2002, Yuan:2022}.
To analyze our data for a Drude-type contribution in the imaginary part of the dielectric function, we first compute the pure intraband term by subtracting the dielectric function at 0 K lattice temperature without any disorder from that for an electron, lattice (ZG), and magnetic temperature of 300\,K.
This difference (orange solid line in Fig.\,\ref{fig:drude}) corresponds to phonon-assisted and magnon-assisted optical transitions.
Clearly, Fig.\ \ref{fig:drude} shows Drude-like behavior of our data between 0.2 and 0.9 eV.
Fitting to Eq.\,\eqref{eq:drude}, results in $\omega_{\mathrm{p}} = 6.52$\,eV and $\omega_{\mathrm{\tau}} = 0.38$\,eV, compared to $\omega_{\mathrm{p}} = 4.09$\,eV and $\omega_{\mathrm{\tau}} = 0.02$\,eV from the fit to experiment \cite{Ordal:1985}.
The plasma frequencies agree reasonably well and the difference in relaxation frequency may be attributed to sample impurities that would lead to a smaller relaxation frequency obtained from the experiment.
We note that the low-energy range of our results suffers from finite-size effects and does not diverge like the Drude term, which can be corrected by employing larger supercells.
Also, the low photon energy range of the dielectric function requires dense $\mathbf{k}$-point sampling and appropriate choice of broadening parameters \cite{Kohn:1957, Peierls:1974, Trushin:2010}.
Given these convergence limitations, our data indicates that the ZG approach captures Drude behavior in metals even for our supercell sizes, and more quantitative studies, in particular of the predicted lifetimes (relaxation frequencies), are left for future investigations.

\section{\label{sec:cncl}Conclusions}

Temperature-dependent optical spectra and first-order magneto-optical spectra of ferromagnetic BCC Fe are predicted based on the supercell approach and Williams-Lax theory for lattice and magnetic disorders.
These effects cannot be described by simulations using the ground state primitive cell and we show that temperature-dependent spectra agree better with experimental spectra.
The calculated optical spectra capture the phonon- and magnon-assisted intraband transitions and show a red shift of the dominant peak near 2.7\,eV that was reported before from the experiment.
Our results for temperature-dependent spectra indicate that this redshift originates predominantly from the magnetic temperature which causes thermal demagnetization and a resulting reduction of exchange splitting.
The supercell approach used here to simulate thermal disorder also describes phonon and magnon-assisted intraband transitions that affect the spectrum at photon energies below 2\,eV.
We show that these significant contributions to the spectra are due to increased optical matrix elements in the presence of disorder and discuss the connection to the Drude model.
Lastly, our data shows band anomalies at finite temperatures as kinks that are caused by electron-phonon and electron-magnon coupling.
Our study demonstrates the possibility of investigating temperature-dependent optical properties of magnetic materials through first-principles simulations.
In the future, this method might be expanded to explain the magneto-optical behavior near the critical temperature and advance the optical application of magnetic materials, including further research on different spin textures such as antiferromagnets and non-collinear magnetic structures at finite temperatures.

\begin{acknowledgements}
This work was undertaken as part of the Illinois Materials Research Science and Engineering Center, supported by the National Science Foundation MRSEC program under NSF Award No.\ DMR-1720633.
This work made use of the Illinois Campus Cluster, a computing resource that is operated by the Illinois Campus Cluster Program (ICCP) in conjunction with the National Center for Supercomputing Applications (NCSA) and which is supported by funds from the University of Illinois at Urbana-Champaign.
This research is part of the Blue Waters sustained-petascale computing project, which is supported by the National Science Foundation (awards OCI-0725070 and ACI-1238993) and the state of Illinois.
Blue Waters is a joint effort of the University of Illinois at Urbana-Champaign and its National Center for Supercomputing Applications.
\end{acknowledgements}

\bibliographystyle{apsrev}
\bibliography{./main.bib}

\end{document}


\title{Supplementary Materials for \\ 
Temperature-dependent optical and magneto-optical spectra of ferromagnetic BCC Fe}

\author{Kisung Kang}
\affiliation{ 
The NOMAD Laboratory at the FHI of the Max-Planck-Gesellschaft and IRIS-Adlershof of the Humboldt-Universit\"{a}t zu Berlin, Faradayweg 4-6, 14195 Berlin, Germany
}

\author{David G. Cahill}
\affiliation{Department of Materials Science and Engineering, University of Illinois at Urbana-Champaign, Urbana, IL 61801, USA}
\affiliation{Department of Physics, University of Illinois at Urbana-Champaign, Urbana, IL 61801, USA}
\affiliation{Materials Research Laboratory, University of Illinois at Urbana-Champaign, Urbana, IL 61801, USA}

\author{Andr\'e Schleife}
\email{schleife@illinois.edu}
\affiliation{Department of Materials Science and Engineering, University of Illinois at Urbana-Champaign, Urbana, IL 61801, USA}
\affiliation{Materials Research Laboratory, University of Illinois at Urbana-Champaign, Urbana, IL 61801, USA}
\affiliation{National Center for Supercomputing Applications, University of Illinois at Urbana-Champaign, Urbana, IL 61801, USA}

\maketitle

\section{Description of lattice and magnetic interactions}
\begin{figure}[H]
\centering
\includegraphics[width=0.69\columnwidth]{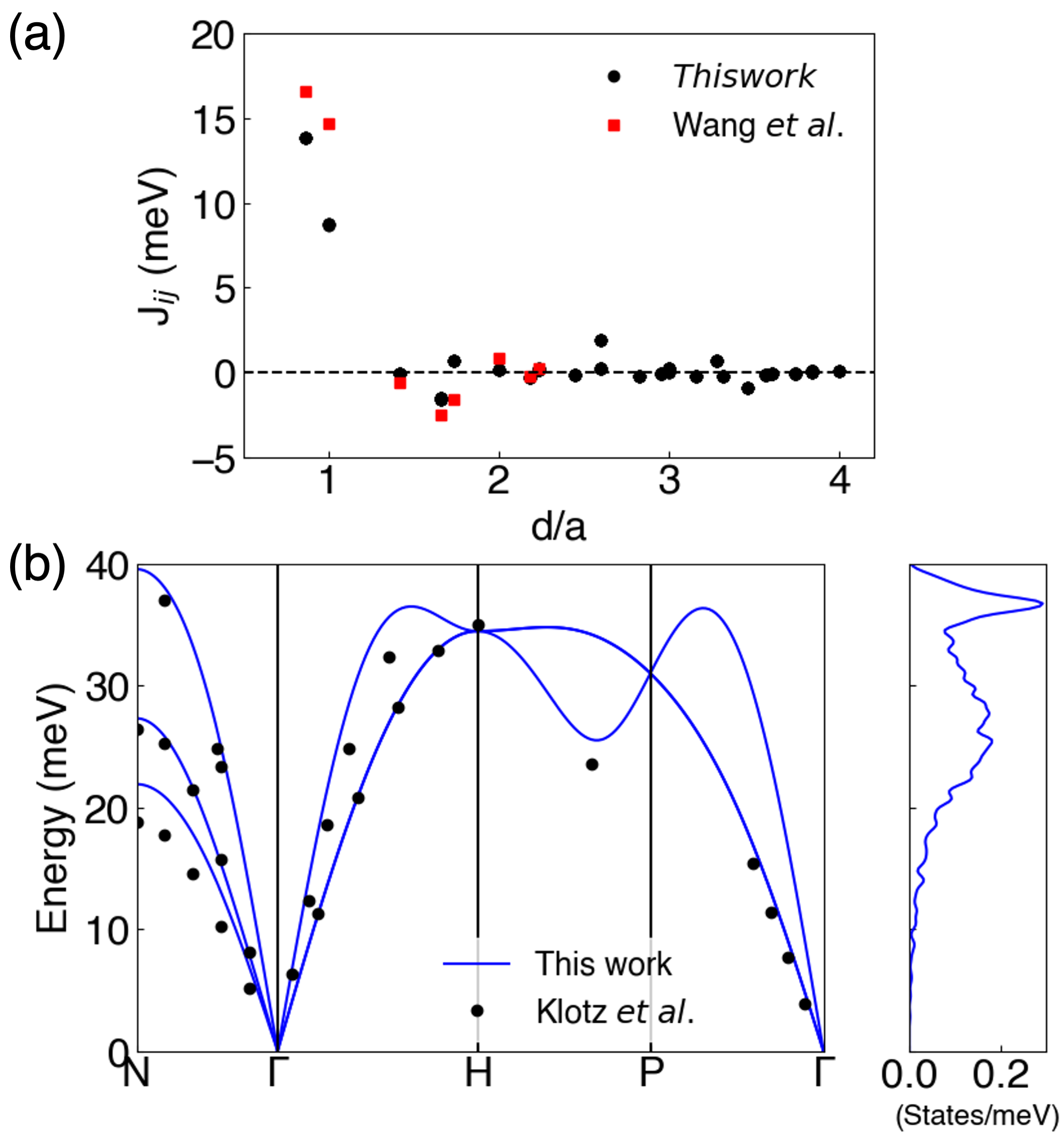}
\caption{\label{fig:interaction}(Color online.)
(a) Exchange coefficients ($J_{ij}$) as a function of relative distance in terms of the lattice parameter $a$ and red square markers are taken from other DFT-GGA study\cite{Wang:2010}, (b) The phonon dispersion curve and density of states of ferromagnetic BCC Fe. Orange triangle markers are signals from the neutron scattering measurement of Ref.\,\onlinecite{Klotz:2000}.
}
\end{figure}

\begin{figure}[H]
\centering
\includegraphics[width=0.59\columnwidth]{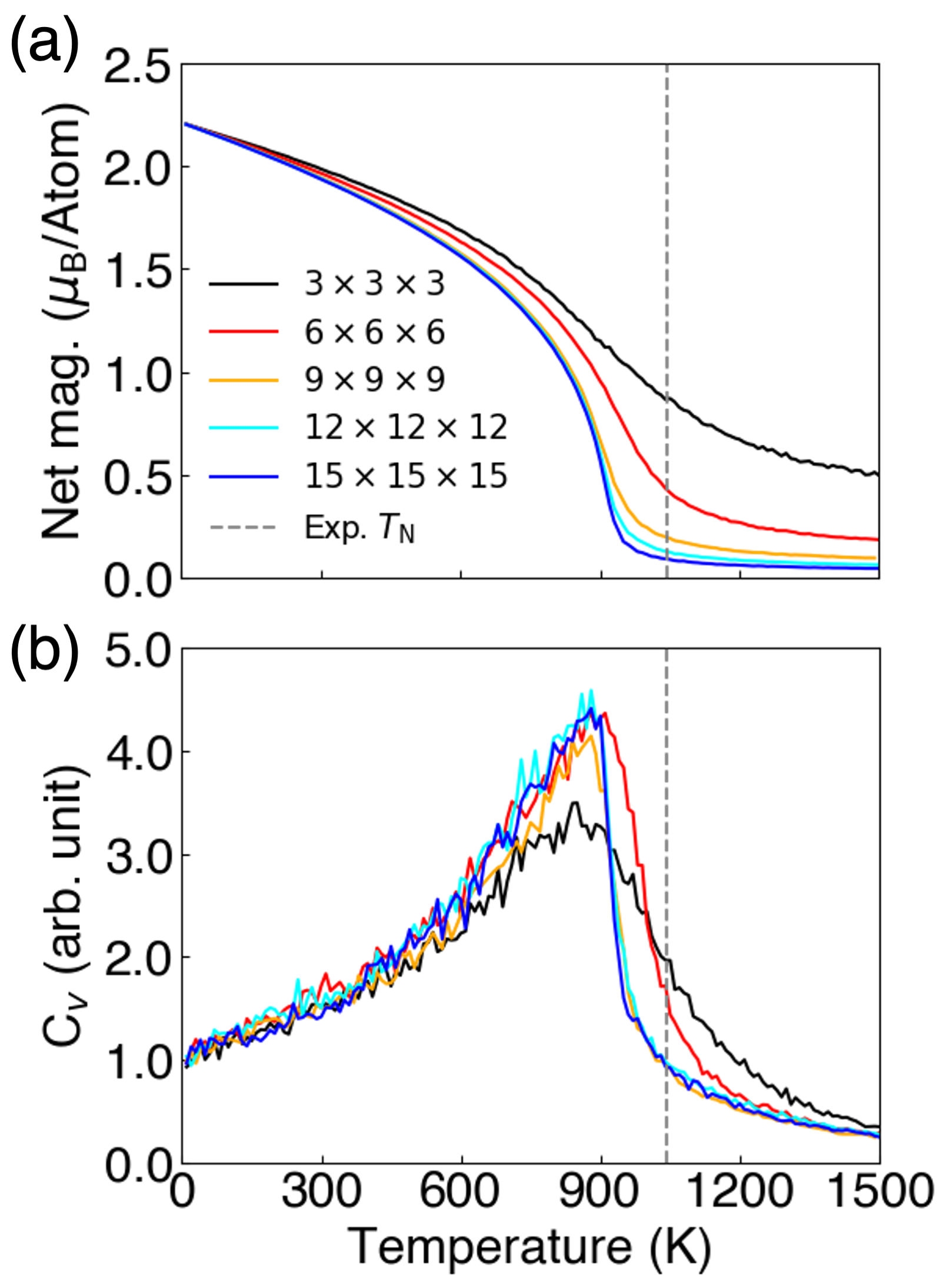}
\caption{\label{fig:thermo}(Color online.)
Thermodynamic observables for the determination of the magnetic phase transition temperature from atomistic spin dynamics simulation with different supercell sizes. (a) $\mathrm{M}_{\mathrm{sub}}$ is a net magnetization per each Fe atom along [001] direction of the conventional cell, and (b) $C_{v}$ is a magnetic heat capacity. The measured Curie temperature (gray dashed line) is from Ref.\ \onlinecite{Ashcroft:1976}}
\end{figure}

To demonstrate our appropriate description of lattice and magnetic interactions in this work, we compare calculated exchange coefficients from the magnetic structure and phonon dispersion of ferromagnetic BCC Fe with other references\cite{Wang:2010, Klotz:2000}.
Calculated exchange coefficient in Fig.\,\ref{fig:interaction} (a) are plotted up to the relative distance ($d/a$) of 4.
Compared to Wang's DFT-GGA study\cite{Wang:2010}, it shows good agreement in general but the coefficients of first and second neighbors are somewhat underestimated, leading to the slightly lower Curie temperature in Fig.\,\ref{fig:thermo}.
Calculated phonon dispersion also matches well with the signals from the neutron scattering measurement implemented in Ref.\,\onlinecite{Klotz:2000} as shown in Fig\,\ref{fig:interaction} (b).

Fig.\,\ref{fig:thermo} exhibits the convergence behavior of temperature-dependent net magnetization per each Fe atom along [001] direction of the conventional cell and magnetic heat capacity from atomistic spin dynamics in terms of the simulation cell size, concluding the selection of the $9\times9\times9$ supercell.

\newpage

\section{The thermal effect of the electron temperatures on optical spectra}
The thermal effect of the electron temperatures on the optical spectra is negligible since the occupation number of electron band structure only changes a few meV near the Fermi level below 300\,K.

\begin{figure}[H]
\centering
\includegraphics[width=0.59\columnwidth]{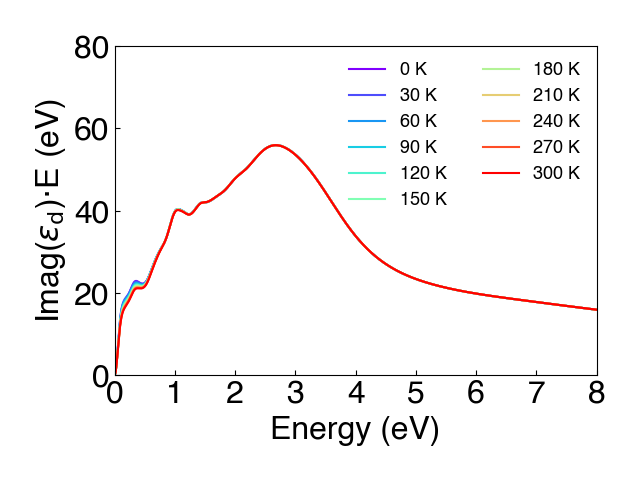}
\caption{\label{fig:imag-elec}(Color online.)
The diagonal components of the temperature-dependent imaginary optical spectra, $(\varepsilon_{\mathrm{d}}-1) \cdot E$, of ferromagnetic BCC Fe from DFT results with electron temperatures from 0\,K to 300\,K.
}
\end{figure}

\newpage

\section{The impact of the electron lifetime}
\begin{figure}[b]
\centering
\includegraphics[width=0.59\columnwidth]{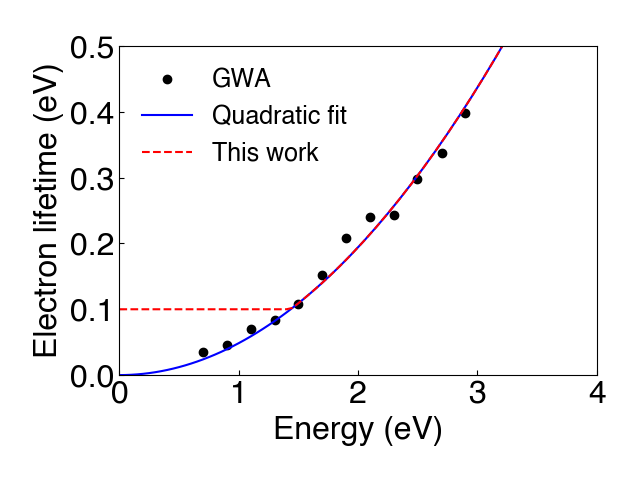}
\caption{\label{fig:gw}(Color online.)
The averaged electron lifetime of spin majority and minority states from the GW approach implemented by Zhukov \emph{et al.}\cite{Zhukov:2006} (Black circle markers).
The blue solid line represents the quadratic fit (quadratic coefficient: 0.0485) of GW results\cite{Zhukov:2006}, while the red dashed line displays the actual lifetime exploited in all figures in the manuscript.
}
\end{figure}

As described in the main text, all plots use the frequency-dependent lifetime.
The quadratic frequency dependence of electron lifetime plotted in Fig.\,\ref{fig:gw} is fitted using electron-electron scattering predicted from the $GW$ approach\cite{Zhukov:2006}.
Here, we use the averaged lifetime over the spin majority and minority results in Zhukov's work\cite{Zhukov:2006}.
The averaged electron-phonon scattering below 1\,eV from the study of Carva \emph{et al.}\cite{Carva:2013} is about 0.02\,eV, which is a too small value to provide enough smearing on the optical spectra and to prevent the spiky peaks caused by the sparse k-point sampling near Fermi surface.
Thus, here we select the lifetime of 0.1\,eV at a low energy range plotted as a dashed red line in Fig.\,\ref{fig:gw}.

\begin{figure}[H]
\centering
\includegraphics[width=0.63\columnwidth]{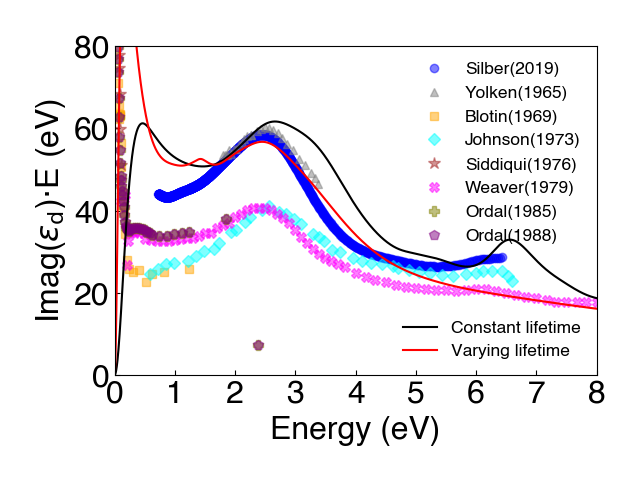}
\caption{\label{fig:diag-const}(Color online.)
The imaginary parts with a constant electron lifetime of $\Gamma=0.3$\,eV (black solid line) and a frequency-dependent lifetime (red solid line) of the averaged diagonal components of the optical spectra, $(\varepsilon_{\mathrm{d}}-1) \cdot E$, of ferromagnetic BCC Fe from DFT results with the all-included electron, lattice (ZG), and magnetic temperatures of 300\,K.
Markers display measured values at room temperature\cite{Yolken:1965, Blotin:1969, Johnson:1974, Siddiqui:1977, Ordal:1985, Ordal:1988, Silber:2019} and 4.2\,K\cite{Weaver:1979}.
}
\end{figure}

Fig\,\ref{fig:diag-const} exhibits the impact of the selection of electron lifetime.
Constant lifetime (black solid line) applies excessive smearing in the low energy range below 1.0\,eV, leading to a reduction in the intensity of the intraband transition peak and contradicting with the measured spectra\cite{Blotin:1969, Siddiqui:1977, Ordal:1985, Ordal:1988, Weaver:1979}.
In addition, a peak near 2.7 eV suffers a somewhat weak smearing effect, leading to a bit wide peak width.
Such weak smearing in the high energy range can be beneficial since a peak near 6.7 eV appears and matches well with the peak shown in measured spectra\cite{Weaver:1979, Johnson:1974}.
Although varying lifetime (red solid line) misses this peak due to the strong smearing effect at high energy, we choose the frequency-dependent lifetime because of the better description near visible light ranges popularly investigated in experiments thus far.

\newpage

\section{The temperature dependence of real part of optical spectra and first-order magneto-optical spectra}

\begin{figure}[H]
\centering
\includegraphics[width=0.83\columnwidth]{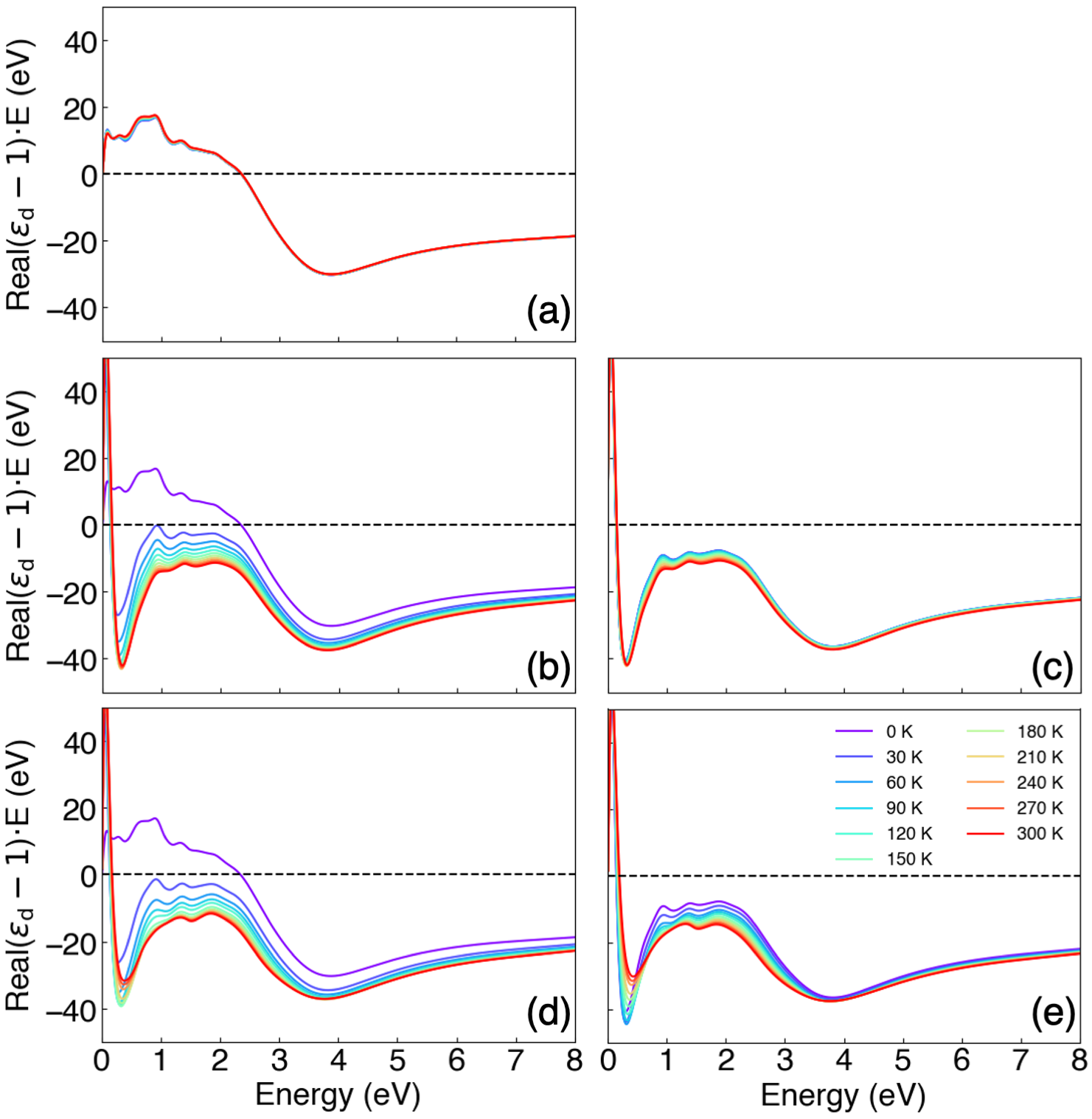}
\caption{\label{fig:diag-real}(Color online.)
The real part of the averaged diagonal components of the temperature-dependent optical spectra, $(\varepsilon_{\mathrm{d}}-1) \cdot E$, of ferromagnetic BCC Fe from DFT results with (a) electron temperatures, (b) lattice temperatures (CS), (c) lattice temperatures (ZG), (d) magnetic temperatures, and (e) electron, lattice (ZG), and magnetic temperatures from 0\,K to 300\,K.
}
\end{figure}

\newpage

The temperature dependence of real optical spectra plotted in Fig.\,\ref{fig:diag-real} behaves similarly to that of imaginary optical spectra.
Electron temperature shown in Fig.\,\ref{fig:diag-real} (a) barely affects the real optical spectra same as in Fig.\,\ref{fig:imag-elec} (a).
The inclusion of lattice and magnetic temperatures corrects the positive signal below the photon energy of 3.0\,eV at 0\,K.
The lattice temperature dependence of real optical spectra is weakened by the zero-point vibrational motion, as illustrated by the distinction between Fig.\,\ref{fig:diag-real}(b) and (c).
Fig.\,\ref{fig:diag-real} (e) indicates that the magnetic temperature dominantly contributes to the temperature dependence of the spectra after the inclusion of the zero-point vibrational motion.

\begin{figure}[H]
\centering
\includegraphics[width=0.59\columnwidth]{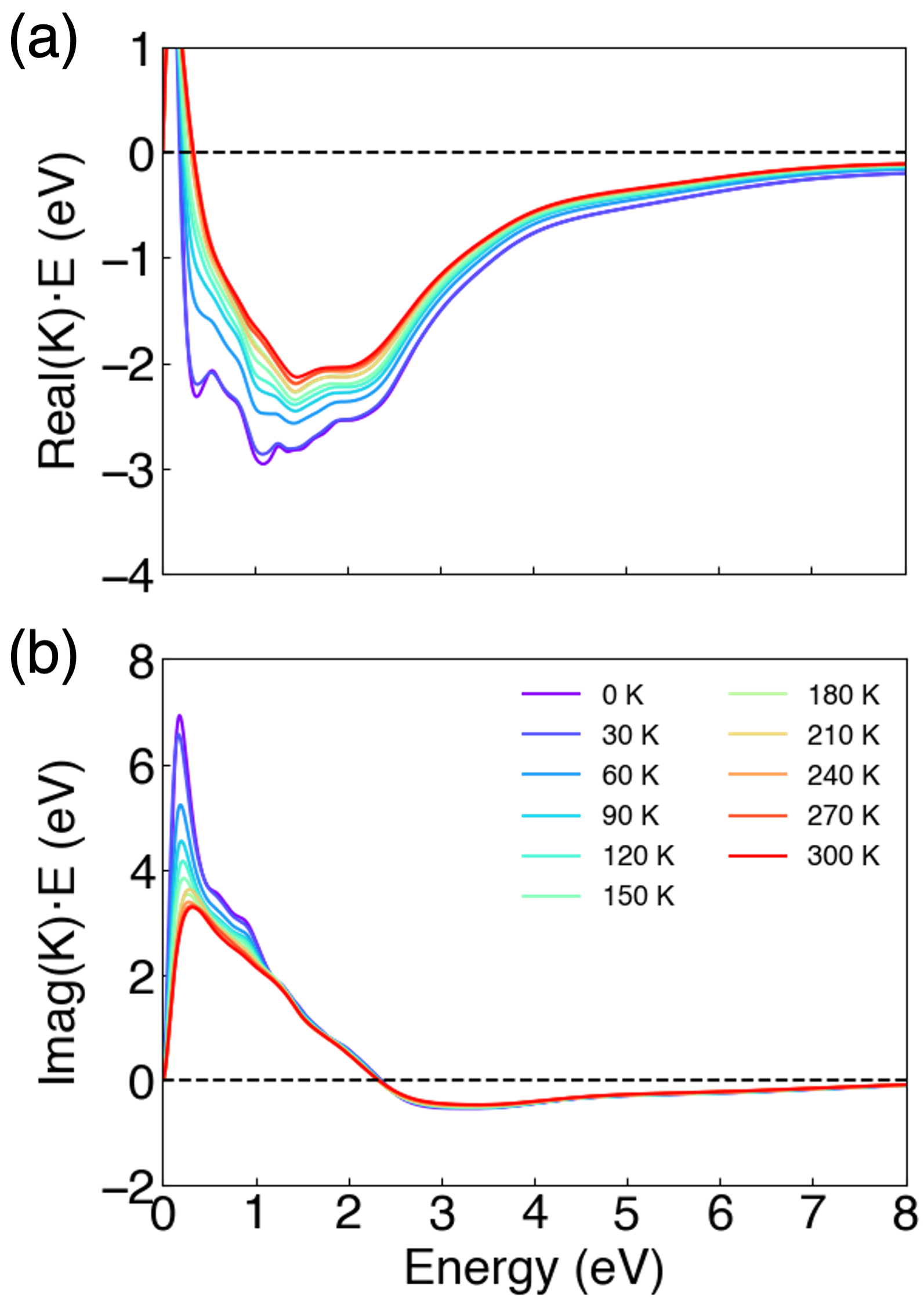}
\caption{\label{fig:first-temp}(Color online.)
(a) Real and (b) imaginary parts of the first-order magneto-optical spectra, $K \cdot E$, of ferromagnetic BCC Fe from DFT results with electron, lattice (ZG), and magnetic temperature from 0\,K to 300\,K.
}
\end{figure}

The temperature dependence of first-order magneto-optical spectra reflects thermal demagnetization which is proportional to magnetic temperature.
As temperature increases, magnetic moments of ferromagnetic BCC Fe start precessing, reducing the net magnetization.
Since the Curie temperature of ferromagnetic BCC Fe is about 1043\,K\cite{Ashcroft:1976}, spectral change up to 300\,K is subtle but the results can describe such a slight reduction of first-order magneto-optical spectrum signal.

\newpage

\section{Second-order magneto-optical spectra}
We studied one of the second-order magneto-optical effects, $G_{s}$, which can be calculated by following the equation for magnetization oriented toward [001]\cite{Hamrlova:2016}.
\begin{equation}
\label{eq:second}
    G_{s}=\varepsilon_{\mathrm{zz}}^{[001]}-\frac{\varepsilon_{\mathrm{xx}}^{[001]}+\varepsilon_{\mathrm{yy}}^{[001]}}{2}
\end{equation}
Here, we emphasize that the magnitude of second-order magneto-optical spectra, $G_{s} \cdot E$, is about ten times smaller than that of first-order magneto-optical spectra, $K \cdot E$ (See Fig.\,3 in the manuscript).

\begin{figure}[H]
\centering
\includegraphics[width=0.59\columnwidth]{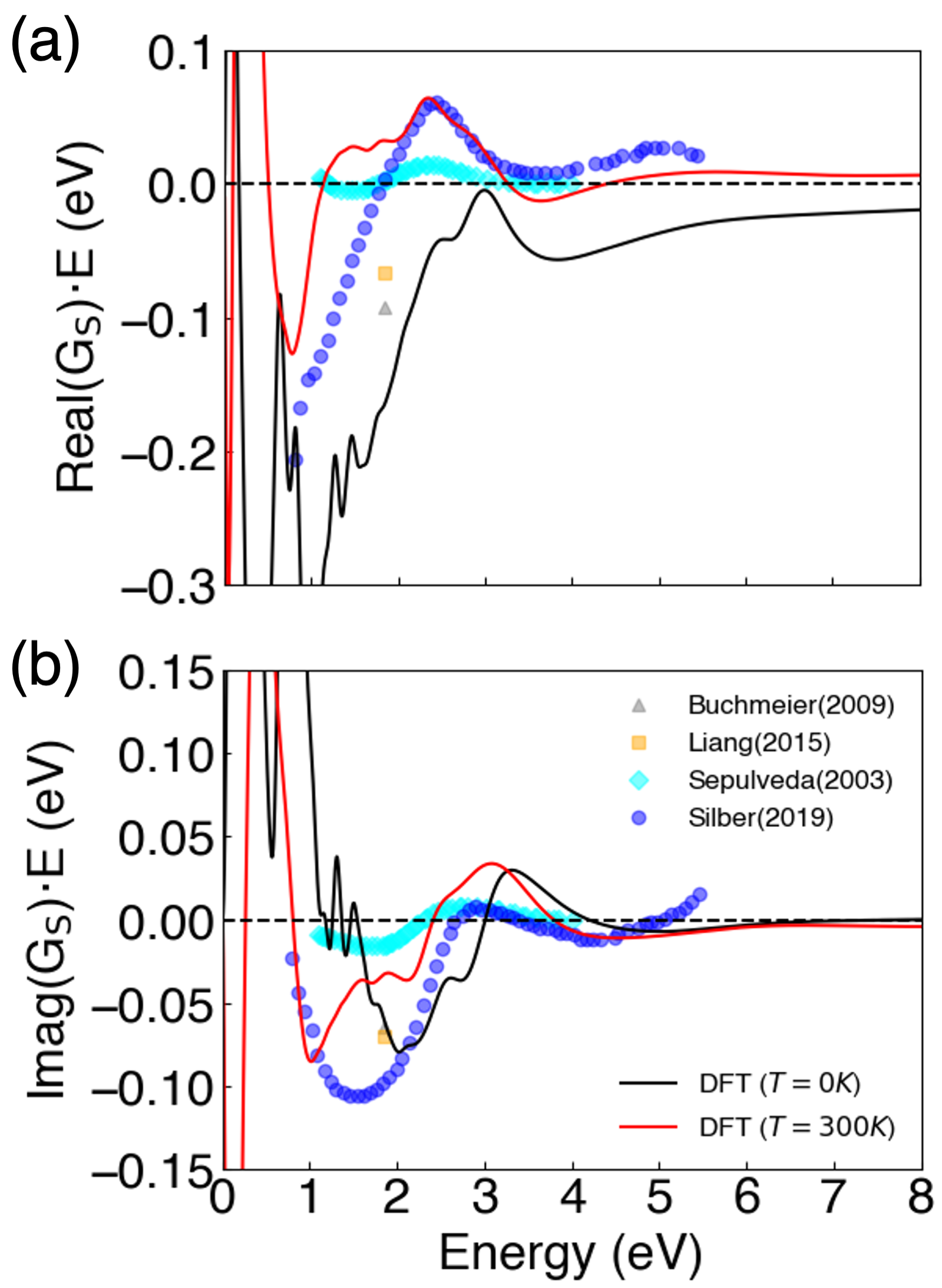}
\caption{\label{fig:second}(Color online.)
(a) Real and (b) imaginary parts of the second-order magneto-optical spectra, $G_{s} \cdot E$, of ferromagnetic BCC Fe from DFT results with 0\,K without zero-point displacement (solid gray line) and electron, lattice (ZG), and magnetic temperature of 300\,K (brown solid line).
Markers display measured values at room temperature.\cite{Sepulveda:2003, Buchmeier:2009, Liang:2015, Silber:2019}.
}
\end{figure}

\begin{figure}[H]
\centering
\includegraphics[width=0.80\columnwidth]{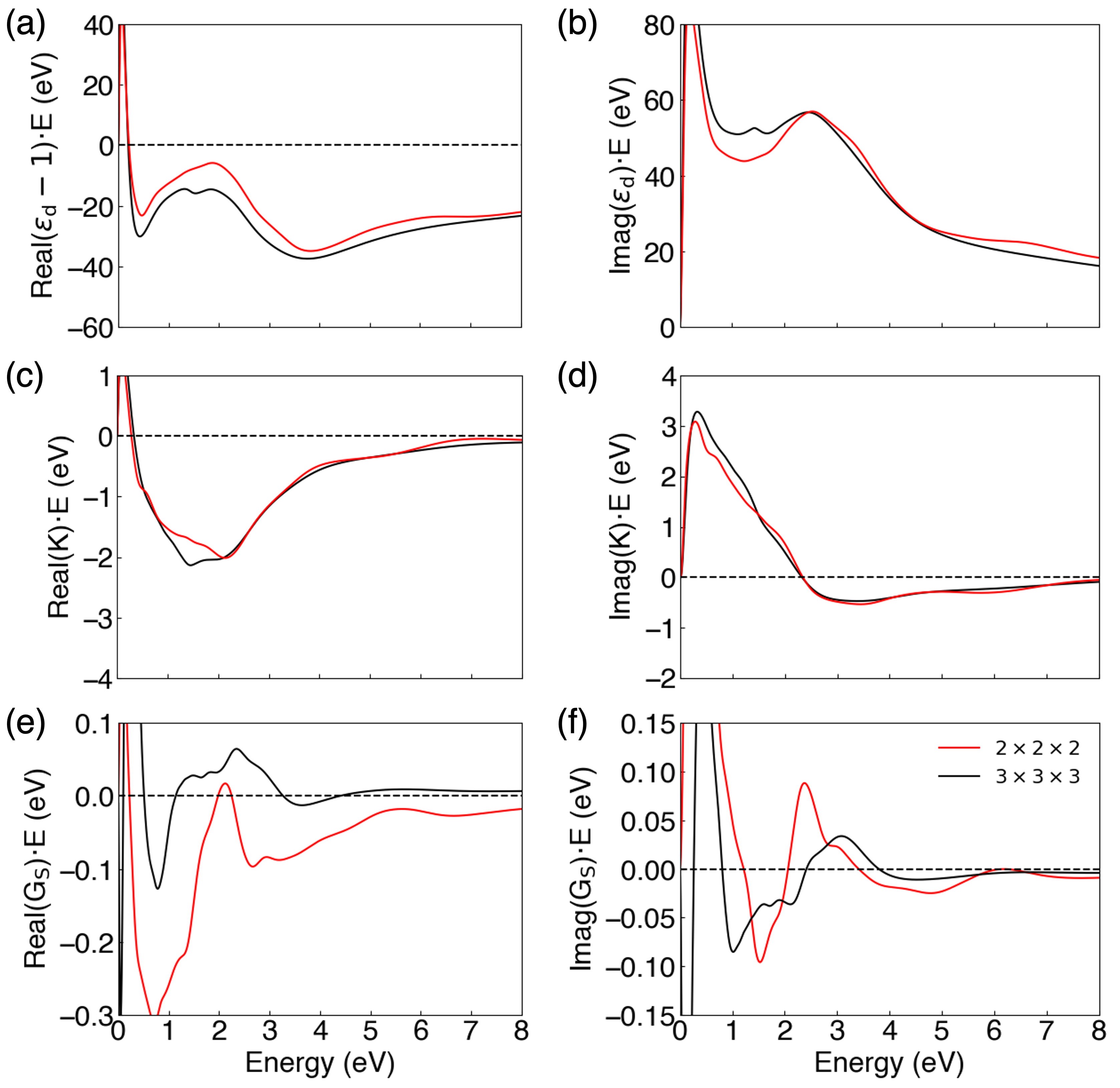}
\caption{\label{fig:super}(Color online.)
Spectra at electron, lattice (ZG), and magnetic temperature of 300\,K with $2\times2\times2$ supercells averaged over 217 snapshots  (Red solid line) and $3\times3\times3$ supercells averaged over 27 snapshots (Black solid line) of (a), (b) optical spectra ($(\varepsilon_{\mathrm{d}}-1) \cdot E$), (c), (d) first-order magneto-optical spectra ($K \cdot E$), and (e), (f) second-order magneto-optical spectra ($G_{s} \cdot E$). 
}
\end{figure}

\begin{figure}[H]
\centering
\includegraphics[width=0.80\columnwidth]{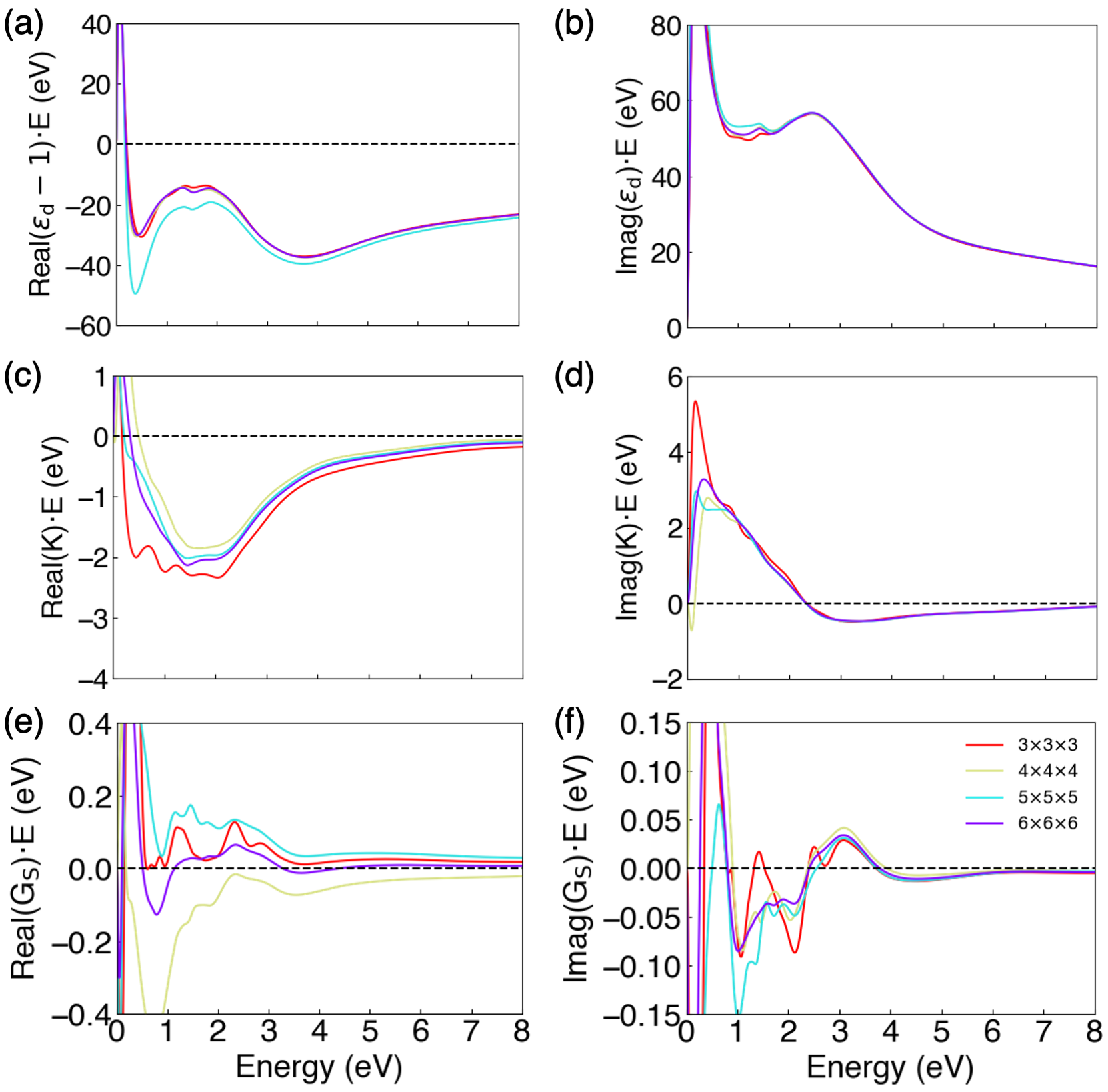}
\caption{\label{fig:kpoint}(Color online.)
Spectra at all-included electron, lattice (ZG), and magnetic temperature of 300\,K with randomly shifted $\Gamma$-centered $3\times3\times3$, $4\times4\times4$, $5\times5\times5$, and $6\times6\times6$ 
$\mathbf{k}$-point grids of (a), (b) optical spectra ($(\varepsilon_{\mathrm{d}}-1) \cdot E$), (c), (d) first-order magneto-optical spectra ($K \cdot E$), and (e), (f) second-order magneto-optical spectra ($G_{s} \cdot E$).
}
\end{figure}

Therefore, the mean absolute error is considerable compared to the absolute magnitude of second-order spectra as shown in Fig.\,\ref{fig:error}.
It might require more disorder snapshots, sufficiently large supercells (Fig.\,\ref{fig:super}), or very dense $\mathbf{k}$-point grids (Fig.\,\ref{fig:kpoint}) to get accurate spectra.
Thus, here we only discuss the qualitative analysis of second-order magneto-optical spectra, which requires a careful interpretation.
For the real part, a peak near 2.5\,eV shown in the measured spectrum from Silber \emph{et al.}\cite{Silber:2019} is captured in the calculated spectrum at an all-included temperature of 300\,K.
The imaginary part of second-order magneto-optical spectra presents a valley shape near 1.6\,eV and a peak near 2.8\,eV plotted in Silber's spectrum\cite{Silber:2019} and calculated second-order signal also contains these features.
Both real and imaginary spectra of second-order magneto-optical effect at the magnetic temperature of 300\,K are redshifted from 0\,K spectra, and this redshifting goes beyond the measurement results.

\newpage

\section{Unfolded electronic band structure at electron temperature and lattice (CS) temperature}
At an electron temperature of 300\,K, it only changes the occupation number of the electronic band structure and thus the unfolded band structure is almost identical to the spin-polarized band structure of the primitive cell shown in Fig.\,5 (a) on the main text.
Their small difference might originate from the spin-orbit coupling effect which is weak in BCC Fe.
The difference in band structures between two lattice temperatures (CS) in Fig.\,S9 (b)\cite{Supplement} and (ZG) in Fig.\,5 (b) is negligible and the characteristic band kinks exhibit nearly the same features because the impact of zero-point vibrational motion at 300\,K, becomes comparable as discussed in the peak redshifting.
Unfolded band structure at a lattice (CS) temperature shows only slightly stronger renormalization over $\mathbf{k}$-space compared to that at a lattice (ZG) temperature shown in Fig.\,5 (b) on the main text.

\begin{figure}[H]
\centering
\includegraphics[width=0.98\columnwidth]{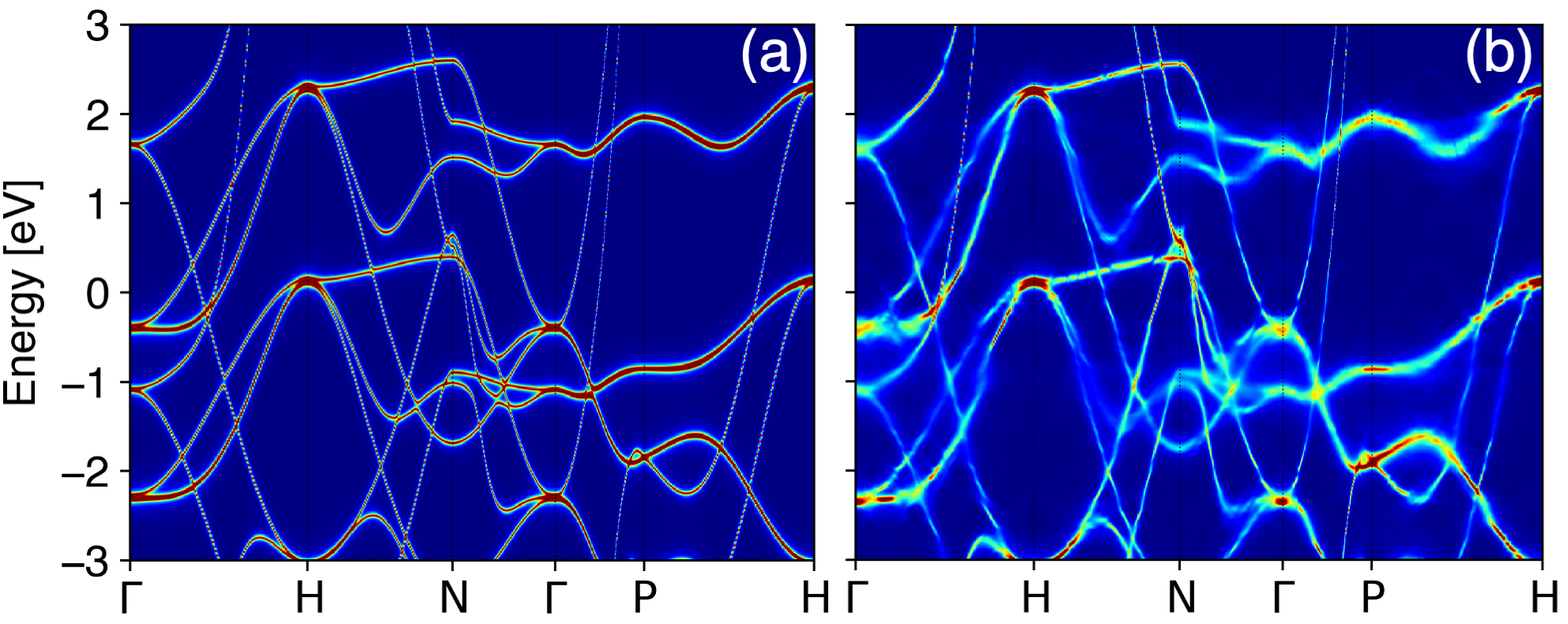}
\caption{\label{fig:band-elec}(Color online.)
Unfolded electronic band structure of ferromagnetic BCC Fe averaged over supercell configurations at (a) electron temperature and (b) lattice (CS) temperature of 300\,K.
}
\end{figure}

\newpage

\section{The population distribution of optical matrix transitions}
When we count the optical transition matrix elements with all $p^{2}_{x}$, their populations show an almost identical distribution in terms of the energy difference between conduction and valence bands regardless of temperature.
However, when we consider the matrix elements which can sufficiently contribute to the optical spectra ($p_{x}^{2}>0.05$) in Fig.\,7 (b) on the main text, the population difference between different temperatures in the low photon energy range clearly elucidates that the change of matrix elements dominantly contributes to the magnon-assisted intraband transitions.

\begin{figure}[H]
\centering
\includegraphics[width=0.50\columnwidth]{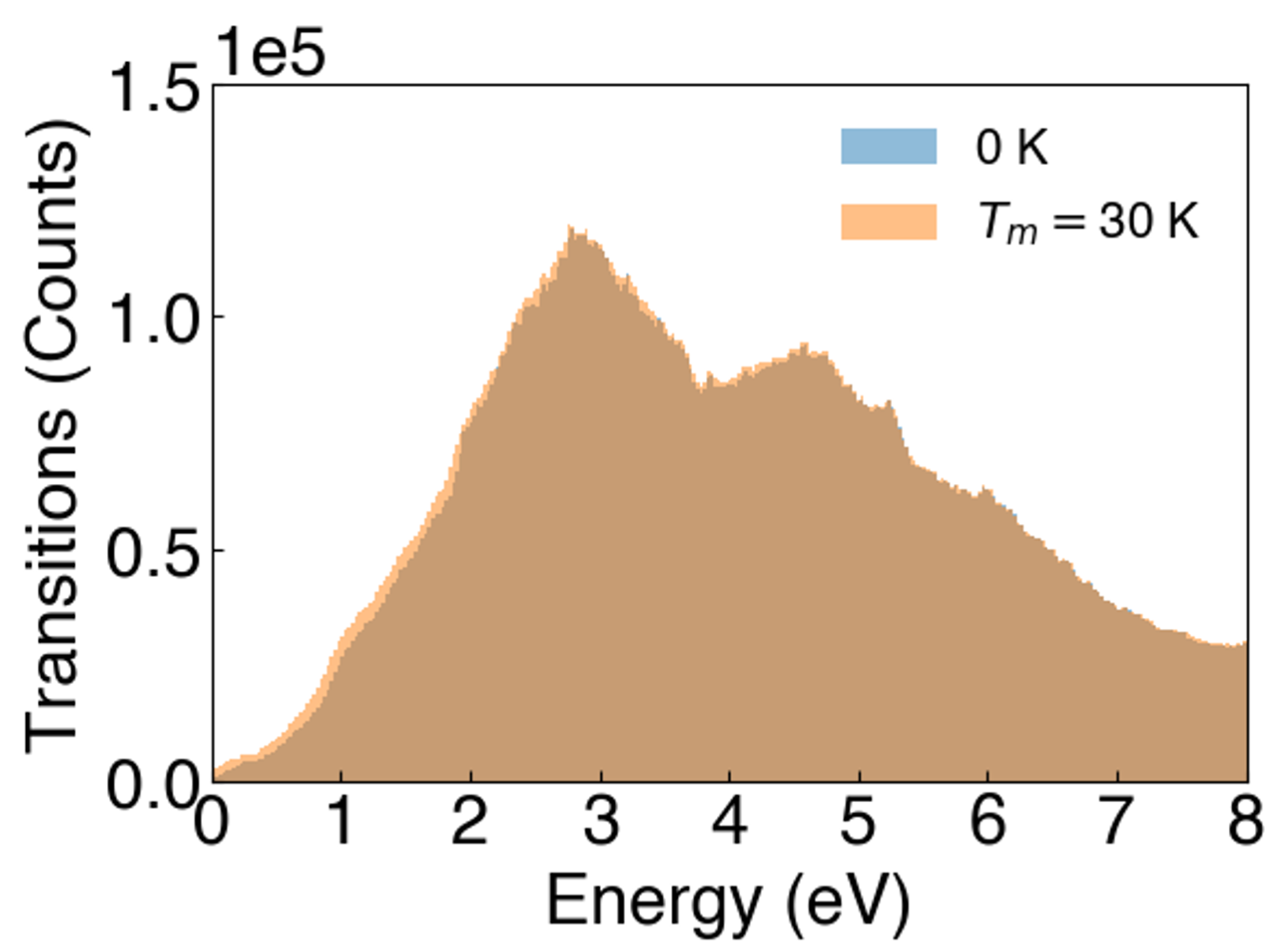}
\caption{\label{fig:matrix-all}(Color online.)
The population distribution of the optical transition matrix elements with all $p_{x}^{2}$ in terms of the energy difference between conduction and valence bands at 0K without zero-point displacement (cyan bars) and magnetic temperature of 30K (orange bars).
}
\end{figure}

\newpage

\section{Convergence check of lattice and magnetic disorders with different supercell sizes}
\begin{figure}[H]
\centering
\includegraphics[width=0.45\columnwidth]{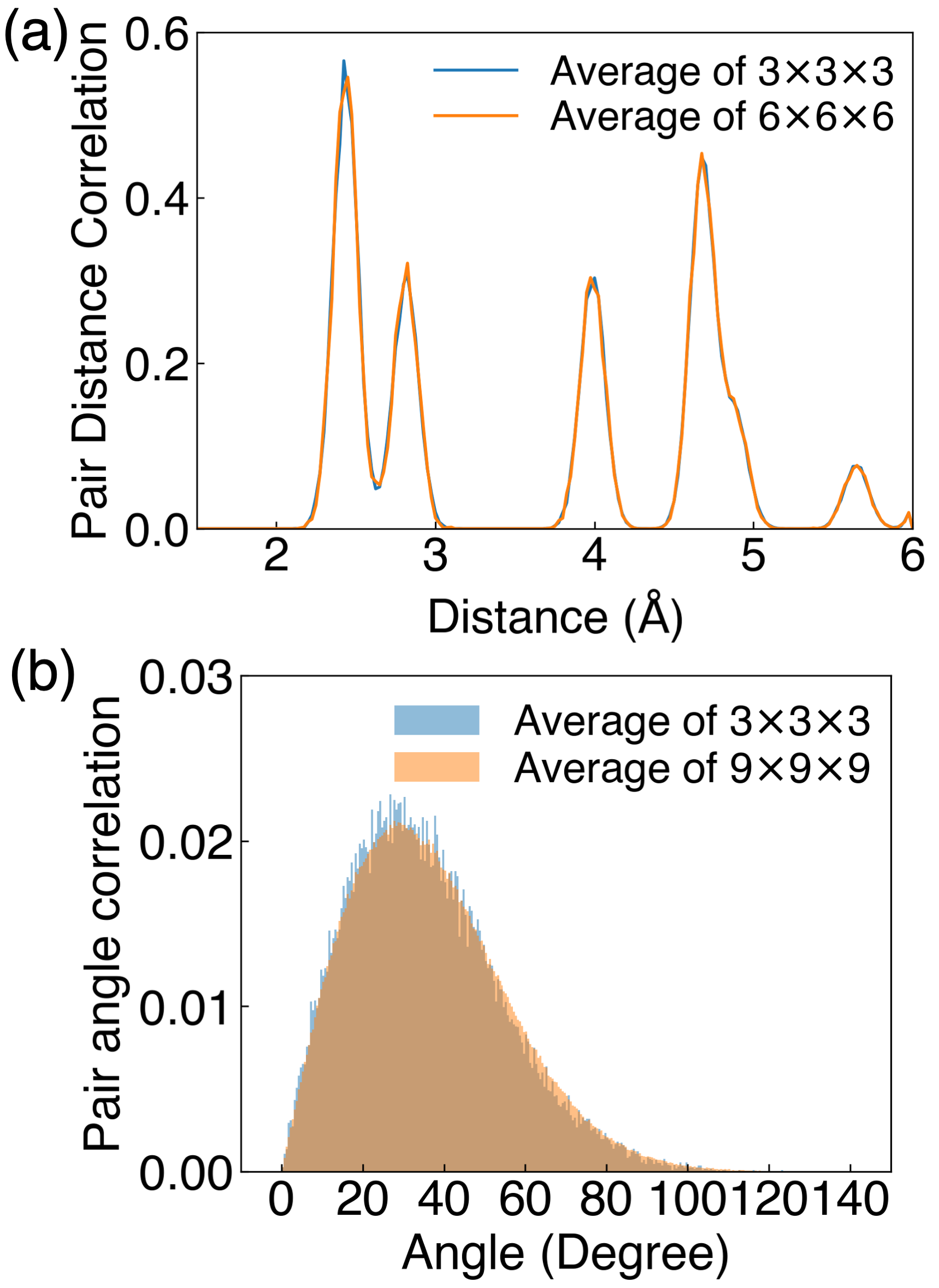}
\caption{\label{fig:size}(Color online.)
(a) The averaged pair distribution function densities of the atomic structure from 27 snapshots of $3\times3\times3$ supercells and 8 snapshots of $6\times6\times6$ supercells. (b) The averaged pair angle distribution function densities of the atomic structure from 27 snapshots of $3\times3\times3$ supercells and 1 snapshots of $9\times9\times9$ supercells.
}
\end{figure}

For a reliable description of lattice and magnetic disorders, it is crucial to converge the distribution of atomic displacements and magnetic fluctuations in terms of supercell sizes.
We examined various supercells to assess their convergence, as illustrated in Fig.\,\ref{fig:size}.
In Fig.\,\ref{fig:size} (a), the averaged pair distribution function densities of the atomic structure were compared between 27 snapshots of $3\times3\times3$ supercells and 8 snapshots of $6\times6\times6$ supercells, yielding a mean absolute error of 0.004.
Similarly, in Fig.\,\ref{fig:size} (b), the averaged pair angle distribution function densities of the magnetic structure were compared between 27 snapshots of $3\times3\times3$ supercells and 1 snapshot of a $9\times9\times9$ supercell, resulting in a mean absolute error of 0.0005.
The pair angle distribution function illustrates the angle difference distribution between two distinct magnetic sites within the magnetic structure.

\newpage

\section{Standard deviation of (magneto)-optical spectra over the snapshots}
\begin{figure}[H]
\centering
\includegraphics[width=0.80\columnwidth]{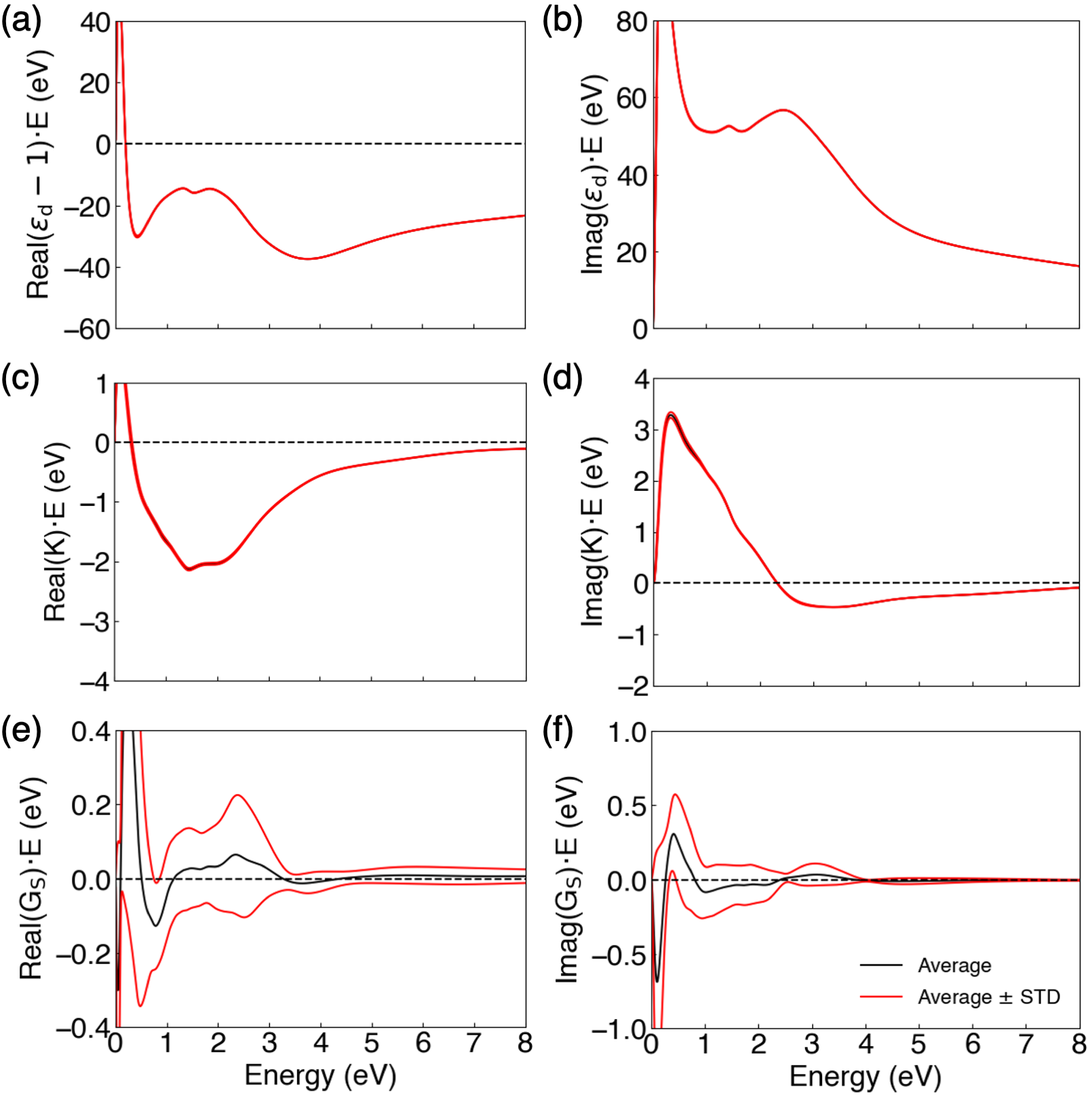}
\caption{\label{fig:error}(Color online.)
Average and average with the standard deviation (STD) of optical spectra ($(\varepsilon_{\mathrm{d}}-1) \cdot E$), first-order magneto-optical spectra ($K \cdot E$), and second-order magneto-optical spectra ($G_{s} \cdot E$) from the ensemble at electron, lattice (ZG), and magnetic temperature of 300\,K.
}
\end{figure}

The optical signal is about ten times greater than the first-order magneto-optical spectra, which, in turn, is roughly ten times larger than the second-order magneto-optical spectra.
Since their standard deviations are comparable, their relative errors are significantly different as shown in Fig.\,\ref{fig:error}.
Thus, the errors of optical conductivity and first-order magneto-optical spectra are reliably small, while that of second-order magneto-optical spectra overwhelms the averaged value.

\newpage

\section{Convergence check of (magneto-)optical spectra with different supercell sizes}
The larger supercell size corresponds to the denser $\mathbf{q}$-point sampling in terms of phonon dispersion.
In addition, it implies the lattice waves with longer wavelengths including low-energy acoustic waves  which are significant to describe the thermal properties at low temperatures.
Optical and first-order magneto-optical spectra exhibit reasonably good convergence with some corrections happening at low phonon energy ranges below 3\,eV, while second-order magneto-optical spectra only display qualitative agreement.

\newpage

\section{Covergence check of (magneto-)optical spectra with different $\mathbf{k}$-point grids}
Optical conductivity and first-order magneto-optical spectra exhibit the convergence with respect to $\mathbf{k}$-point grid, while second-order magneto-optical spectra might require denser $\mathbf{k}$-point grid than $6\times6\times6$, the densest grid that we tested.

\bibliographystyle{apsrev}
\bibliography{./main.bib}